%% file: paper.tex
\documentclass[sigplan, nonacm, screen]{acmart}
\pdfoutput=1
\setcopyright{rightsretained}
\acmPrice{}
\acmDOI{10.1145/3385412.3385992}
\acmYear{2020}
\copyrightyear{2020}
\acmISBN{978-1-4503-7613-6/20/06}
\acmConference[PLDI '20]{Proceedings of the 41st ACM SIGPLAN International Conference on Programming Language Design and Implementation}{June 15--20, 2020}{London, UK}
\acmBooktitle{Proceedings of the 41st ACM SIGPLAN International Conference on Programming Language Design and Implementation (PLDI '20), June 15--20, 2020, London, UK}

\usepackage[T1]{fontenc}
\usepackage{microtype}
\usepackage{booktabs}   
\usepackage{subcaption} 
\usepackage{amsmath}
\usepackage{semantic}
\usepackage{cleveref}
\usepackage{multirow}
\usepackage{listings}
\usepackage{pgfplots}
\pgfplotsset{compat=newest}
\usepackage{subcaption}
\usepackage{xcolor}
\usepackage{lstcoq}
\usepackage{array}
\usepackage{float}
\usepackage{wrapfig}
\usepackage{flushend}

\input{commands}

\title{LL(1) Parsing with Derivatives and Zippers}

\author{Romain Edelmann}
\affiliation{
  \institution{IC, EPFL}           
  \streetaddress{Station 14}
  \city{Lausanne}
  \state{Vaud}
  \postcode{1015}
  \country{Switzerland}                   
}
\email{romain.edelmann@epfl.ch}         

\author{Jad Hamza}
\affiliation{
  \institution{IC, EPFL}           
  \streetaddress{Station 14}
  \city{Lausanne}
  \state{Vaud}
  \postcode{1015}
  \country{Switzerland}                   
}
\email{jad.hamza@epfl.ch}         

\author{Viktor Kun\v{c}ak}
\affiliation{
  \institution{IC, EPFL}           
  \streetaddress{Station 14}
  \city{Lausanne}
  \state{Vaud}
  \postcode{1015}
  \country{Switzerland}                   
}
\email{viktor.kuncak@epfl.ch}         

\begin{document}

\begin{abstract}
In this paper, we present an efficient, functional, and formally verified parsing algorithm for LL(1) context-free expressions based on the concept of derivatives of formal languages.
Parsing with derivatives is an elegant parsing technique, which, in the general case, suffers from cubic worst-case time complexity and slow performance in practice.
We specialise the parsing with derivatives algorithm to LL(1) context-free expressions, where alternatives can be chosen given a single token of lookahead.
We formalise the notion of LL(1) expressions and show how to efficiently check the LL(1) property.
Next, we present a novel linear-time parsing with derivatives algorithm for LL(1) expressions operating on a zipper-inspired data structure.
We prove the algorithm correct in Coq and present an implementation as a part of Scallion, a parser combinators framework in Scala
with enumeration and pretty printing capabilities.
\end{abstract}



\maketitle

\sloppy

\input{introduction.tex}

\input{example.tex}

\input{formalism.tex}

\input{properties.tex}

\input{derivative.tex}

\input{zippy.tex}

\input{coq-proofs.tex}

\input{table-performance.tex}

\input{implementation.tex}

\input{evaluation.tex}

\input{related.tex}

\input{acknowledgements.tex}

\bibliography{paper}

\end{document}

%% file: commands.tex
\newcommand{\PreserveBackslash}[1]{\let\temp=\\#1\let\\=\temp}
\newcolumntype{C}[1]{>{\PreserveBackslash\centering}p{#1}}

\newtheorem*{remark}{Remark}

\DeclareRobustCommand{\values}[0]{\ensuremath{\mathcal{V}}}
\DeclareRobustCommand{\types}[0]{\ensuremath{\mathcal{T}}}
\DeclareRobustCommand{\hastype}[2]{\ensuremath{#1 : #2}}
\DeclareRobustCommand{\fun}[2]{\ensuremath{#1 \to #2}}

\DeclareRobustCommand{\tokentype}[0]{\ensuremath{\texttt{Token}}}
\DeclareRobustCommand{\kinds}[0]{\ensuremath{\mathcal{K}}}

\DeclareRobustCommand{\vars}[1]{\ensuremath{\Sigma_{#1}}}

\DeclareRobustCommand{\syntax}[1]{\ensuremath{\mathcal{S}_{#1}}}

\DeclareRobustCommand{\eps}[1]{\varepsilon_{#1}}
\DeclareRobustCommand{\fail}[0]{\bot}
\DeclareRobustCommand{\elem}[1]{\textit{elem}_{#1}}
\DeclareRobustCommand{\disj}[2]{{#1} \vee {#2}}
\DeclareRobustCommand{\seq}[2]{{#1} \cdot {#2}}
\DeclareRobustCommand{\map}[2]{#1 \circledcirc #2}

\DeclareRobustCommand{\var}[1]{\textit{var}_{#1}}

\DeclareRobustCommand{\getkind}[1]{\texttt{getKind}(#1)}
\DeclareRobustCommand{\getdef}[1]{\texttt{getDef}(#1)}

\DeclareRobustCommand{\matches}[3]{#1 \vdash #2 \leadsto #3}

\DeclareRobustCommand{\prod}[1]{\ensuremath{\textsc{productive}(#1)}}

\DeclareRobustCommand{\nullwith}[2]{\ensuremath{\textsc{nullable}(#1, #2)}}
\DeclareRobustCommand{\nullfun}[1]{\ensuremath{\texttt{nullable}(#1)}}

\DeclareRobustCommand{\first}[1]{\ensuremath{\textsc{first}(#1)}}

\DeclareRobustCommand{\snf}[1]{\ensuremath{\textsc{sn-follow}(#1)}}

\DeclareRobustCommand{\conflict}[1]{\ensuremath{\textsc{has-conflict}(#1)}}

\DeclareRobustCommand{\visitable}[1]{\ensuremath{\textsc{visitable}(#1)}}

\DeclareRobustCommand{\derive}[2]{\ensuremath{\delta_{#1}(#2)}}

\DeclareRobustCommand{\parse}[2]{\ensuremath{\texttt{sParse}(#1, #2)}}

\DeclareRobustCommand{\layers}[2]{\ensuremath{\mathcal{L}_{#2}^{#1}}}
\DeclareRobustCommand{\apply}[1]{\ensuremath{\texttt{apply}(#1)}}
\DeclareRobustCommand{\prepend}[1]{\ensuremath{\texttt{prepend}(#1)}}
\DeclareRobustCommand{\followby}[1]{\ensuremath{\texttt{follow-by}(#1)}}

\DeclareRobustCommand{\context}[2]{\ensuremath{\mathcal{C}_{#2}^{#1}}}

\DeclareRobustCommand{\focustype}[1]{\ensuremath{\mathcal{F}_{#1}}}
\DeclareRobustCommand{\focus}[1]{\ensuremath{\texttt{focus}(#1)}}
\DeclareRobustCommand{\unfocus}[1]{\ensuremath{\texttt{unfocus}(#1)}}
\DeclareRobustCommand{\locate}[2]{\ensuremath{\texttt{locate}(#1, #2)}}
\DeclareRobustCommand{\pierce}[3]{\ensuremath{\texttt{pierce}(#1, #2, #3)}}
\DeclareRobustCommand{\plug}[2]{\ensuremath{\texttt{plug}(#1, #2)}}
\DeclareRobustCommand{\fderive}[2]{\ensuremath{\texttt{derive}(#1, #2)}}
\DeclareRobustCommand{\fnullable}[1]{\ensuremath{\texttt{result}(#1)}}
\DeclareRobustCommand{\fparse}[2]{\ensuremath{\texttt{parse}(#1, #2)}}

\DeclareRobustCommand{\nil}[0]{{\langle}{\rangle}}
\DeclareRobustCommand{\singleton}[1]{{\langle}{#1}{\rangle}}
\DeclareRobustCommand{\app}[2]{{#1}\ {++}\ {#2}}
\DeclareRobustCommand{\cons}[2]{#1\ {::}\ #2}
\DeclareRobustCommand{\sequence}[1]{{\langle}#1{\rangle}}

\DeclareRobustCommand{\none}[0]{\texttt{none}}
\DeclareRobustCommand{\some}[1]{\texttt{some}(#1)}

%% file: introduction.tex

\section{Introduction}

In this paper, we propose a formally verified parsing
approach for LL(1) languages based on derivatives. We
present an implementation of the approach as a parsing
combinator framework, which supports static checks that the grammar is LL(1), and
provides not only parsing and semantic actions, but also enumeration and
pretty-printing functionality. Our implementation remains functional
yet efficient, which allows us to obtain a proof that closely follows implementation.

Whereas parsing is a well understood problem, recent years have seen a renewed interest in approaches
that handle not just language recognition but also syntax tree construction, and that are proven correct formally.
Such parsing techniques can then be leveraged to more productively construct efficient front ends
for verified compilers such as CompCert~\cite{leroy2009formal} and CakeML~\cite{kumar2014cakeml}.
Safe and correct parsers are also crucial for building serialization and deserialization layers
of communication infrastructure, which has been a major target of high-impact security exploits
\cite{CloudflareBug}.

Parsing traditionally uses context-free grammars as the starting specification
formalism and proceeds using table and stack-based algorithms.
Popular techniques include LR parsing~\cite{knuth1965translation, deremer1969practical, lang1974deterministic}, LL parsing techniques~\cite{Lewis:1968:ST:321466.321477,scott2010gll}, recursive descent~\cite{burge1975recursive}, Earley's algorithm~\cite{earley1970efficient}, and the Cocke-Younger-Kasami (CYK) algorithm~\cite{cocke1969programming, kasami1966efficient, younger1967recognition}.
Due to the significant gap between implementation and specification in such approaches, the resulting proofs are often
based on validation as opposed to proofs for the general case \cite{jourdan2012validating}.

In 1964, Brzozowski introduced the concept of derivatives of regular expressions~\cite{brzozowski1964derivatives}.
This concept has proven successful
in many formal proofs of parsing regular expressions and their generalisations~\cite{Pierce:SF1, Posix-Lexing-AFP, MSO_Regex_Equivalence-AFP, Formula_Derivatives-AFP}.

Derivatives of \emph{context-free} expressions~\cite{leiss1991towards} generalize derivatives of regular expressions and have recently been used as an alternative principled approach to understanding context-free parsing~\cite{Danielsson:2010:TPC:1863543.1863585, might2011parsing}, avoiding explicit conversion into pushdown automata.
Context-free expressions offer an algebraic view of context-free grammars.
In addition to describing a language, context-free expressions also describe the value associated with each recognised input sequence, which makes integration into real-world parsers more natural.
The concept of context-free expression derivatives was shown to naturally yield a parsing technique aptly named \emph{parsing with derivatives}~\cite{might2011parsing}, which was later proved to have worst-case cubic complexity~\cite{adams2016complexity}.

For integration into verifiable functional infrastructure, a particularly promising interface are 
parsing combinators~\cite{burge1975recursive, wadler1985replace, hutton1992higher, fokker1995functional, hutton1996monadic}.
Parsing combinator frameworks have been proposed for many functional programming languages such as Haskell~\cite{leijen2001parsec} and Scala~\cite{Haoyi19FastParse, scalaparsercombinators}.
Most implementations of parser combinators use recursive descent for parsing, which has exponential worst-case complexity due to backtracking and can encounter stack overflows with deeply nested structures.
Parsing expression grammars (PEGs)~\cite{ford2004parsing} are also popular in parsing combinators and have been formally verified
\cite{koprowski2010trx}.
While PEGs on the surface resemble context-free grammars, they behave very differently and can exhibit unpredictable behaviour~\cite{redziejowski2008some}.
In addition, merging lexical and syntactic analysis (as often done in the context of PEGs) is not helpful for performance in our experience.

In contrast, LL(1) parsing~\cite{Lewis:1968:ST:321466.321477} is restricted to context-free grammars that can be non-ambiguously parsed given a single token of lookahead and runs in time linear in the input size. An appealing aspect of such grammars is that
they can be algorithmically and efficiently analysed to prevent grammar design errors.
In addition, they are known to provide good performance and error localisation~\cite{Aho:2006:CPT:1177220}.
Previous parsing combinator libraries for LL(1) languages either
do not perform LL(1) checks~\cite{swierstra1996deterministic} or impose restrictions on emptiness when parsing
sequences~\cite{Krishnaswami:2019:TAA:3314221.3314625}, beyond those necessary for the definition of LL(1) languages.

By using the methodology of context-free expression derivatives, we can arrive at an efficient implementation of LL(1) parsing combinators, without introducing needless restrictions. 
We further show that, by embracing Huet's zipper~\cite{huet1997zipper, Mcbride01thederivative} data structure, parsing with derivatives on LL(1) languages can be implemented with linear time complexity.

Our parsing approach sits at an interesting point in the design space of parsers.
The approach offers a \emph{parser combinators interface}, which is naturally embeddable in functional programming languages.
The deeply embedded monoidal~\cite{mcbride2008applicative} nature of our combinators makes it possible to programmatically analyse parsing expressions and enables features such as enumeration of recognised sequences and pretty printing of values.
Thanks to this representation, our approach also supports efficiently checking whether a description is LL(1), ensuring \emph{predictable linear-time parsing}.
Thanks to our use of derivatives for parsing, the parser state is \emph{explicitly encoded as an expression}, not an automaton, which eases reasoning and implementation.
Moreover, this representation of the state makes features such as parsing resumption, error reporting and completion straightforward to support.

\subsection*{Contributions}
\begin{itemize}
 
\item
  We formalize context-free expressions (\emph{syntaxes}) with the expressive power of context-free grammars but with an added ability to describe the values associated with recognised inputs.
Using a new definition of should-not-follow sets, we define \emph{LL(1) syntaxes}, where all alternatives can be resolved given a single token of lookahead.
We show how to use propagator networks~\cite{radul2009propagation} to compute, in linear time,
productivity, nullability, first sets, should-not-follow sets as well as the LL(1) property of syntaxes.

\item
We present an algorithm for parsing with derivatives using LL(1) syntaxes.
Compared to traditional parsing, the algorithm works directly at the level of syntaxes, not on a derived push-down automaton.
We show a technique based on Huet's zipper~\cite{huet1997zipper} to make LL(1) parsing with derivatives efficient.
We show that such \emph{zippy} LL(1) parsing runs in time linear in the input.

\item
We present a Coq formalisation (\url{https://github.com/epfl-lara/scallion-proofs}) of syntaxes and prove the correctness
of the zippy LL(1) parsing with derivatives algorithm and its auxiliary functions.
For performing LL(1) checks, we formalise rule-based descriptions from which we can obtain both an inductive predicate and an equivalent propagator network.

\item
We present Scallion (\url{https://github.com/epfl-lara/scallion}), an implementation of syntaxes as a Scala parser combinators framework with a unique set of features,
implementing LL(1) parsing using derivatives and the zipper data structure.
The framework is efficient, provides error reporting, recovery,
enumeration of accepted sequences, as well as pretty printing.
Benchmarking our framework on a JSON syntax suggests performance comparable to existing 
Scala Parser Combinators~\cite{scalaparsercombinators}, while avoiding stack overflows and providing more features.
\end{itemize}

%% file: example.tex

\lstdefinestyle{scala}{%
    language=Scala,%
    xleftmargin=0mm,%
    aboveskip=2mm,%
    belowskip=2mm,%
    fontadjust=true,%
    columns=[c]fixed,%
    keepspaces=true,%
    basewidth={0.58em, 0.53em},%
    tabsize=2,%
    basicstyle=\renewcommand{\baselinestretch}{0.80}\small\ttfamily,%
    commentstyle=\itshape,%
    keywordstyle=\bfseries,%
    mathescape=true,%
    captionpos=b,%
    framerule=0pt,%
    firstnumber=0,%
    numbersep=0,%
    numberstyle=\tiny,%
    mathescape=false,%
}

\begin{figure*}
\begin{lstlisting}[style=scala]
object JSONParser extends Syntaxes[Token, Kind] {

  val boolValue: Syntax[Value] = accept(BoolKind) { case BoolToken(value) => BoolValue(value) }
  // Definition of other simple values in a similar fashion...

  implicit def separator(char: Char): Syntax[Token] = elem(SepKind(char))

  lazy val arrayValue: Syntax[Value] =
    ('[' ~ repsep(jsonValue, ',') ~ ']').map { case _ ~ values ~ _ => ArrayValue(values) }

  lazy val binding: Syntax[(StringValue, Value)] =
    (stringValue ~ ':' ~ jsonValue).map { case key ~ _ ~ value => (key, value) }

  lazy val objectValue: Syntax[Value] =
    ('{' ~ repsep(binding, ',') ~ '}').map { case _ ~ bindings ~ _ => ObjectValue(bindings) }

  lazy val jsonValue: Syntax[Value] = recursive
    { arrayValue | objectValue | boolValue | numberValue | stringValue | nullValue }
}
\end{lstlisting}
\caption{JSON Parser in Scala using Scallion, the parser combinator framework discussed in this paper.}
\label{fig:example}
\end{figure*}

\section{Example}

To give the flavour of our approach, \Cref{fig:example}~presents a parser for JSON using Scallion, our parser combinators framework implemented
in Scala. The sequencing combinator is denoted by infix $\texttt{\textasciitilde}$, while disjunction is denoted by $\texttt{|}.$
The parser runs efficiently, even though it does not rely on code generation:
with our simple hand-written lexer it takes about 40ms to parse 1MB of raw JSON data into a value of type $\texttt{Value}$, half of which is spent lexing. To provide a comparison point, an ANTLR-\emph{generated} JSON parser~\cite{parr2013definitive,DBLP:conf/pldi/ParrF11,ANTLRexamples}
takes around 15ms per 1MB to produce a parse tree (using its own lexer).

As the Scallion framework is embedded in Scala, we can use the Scala REPL to query the parser.
The following snippets show an example REPL interaction with the framework.
We start by checking the LL(1) property for the top-level \texttt{jsonValue} syntax, and
then list what kinds of tokens can start valid sequences using the method \texttt{first}.
\begin{minipage}{\linewidth}
\begin{lstlisting}[style=scala]
scala> jsonValue.isLL1
// true
scala> jsonValue.first
// Set(NullKind, SepKind('['), ...)
\end{lstlisting}
\end{minipage}
When we feed a valid sequence of tokens to the syntax, we obtain, as expected, a JSON value.

\noindent
\begin{minipage}{\linewidth}
\begin{lstlisting}[style=scala]
scala> val tokens = JSONLexer("[1, 2, 3]")
scala> jsonValue(tokens)
// Parsed(ArrayValue(...), ...)
\end{lstlisting}
\end{minipage}

\noindent
When we feed it an invalid sequence, the syntax duly returns a parse error,
indicating the first unrecognised token and providing the residual syntax at the point of error.
\begin{minipage}{\linewidth}
\begin{lstlisting}[style=scala]
scala> val badtokens = JSONLexer("[1, 2 3]")
scala> val UnexpectedToken(token, rest) =
         jsonValue(badTokens)
// token = NumberToken(3)
// rest represents the residual syntax.
\end{lstlisting}
\end{minipage}
We can then query the residual syntax for valid ways to continue the sequence, or even to resume parsing.
\begin{lstlisting}[style=scala]
scala> rest.first
// Set(SepKind(','), SepKind(']'))

scala> rest(JSONLexer(", 3]"))
// Parsed(ArrayValue(...), ...)
\end{lstlisting}

%% file: formalism.tex

\section{Algebraic Framework for Parsing}

In this section, we formalise the notion of \emph{syntaxes} and describe their semantics as a relation between input token sequences and values.

We consider a set of values \values and a set of types \types. For a value $v \in \values$ and a type $T \in \types$, we denote by \hastype{v}{T} the fact that the value $v$ has type $T$.
We denote by $(v_1, v_2) \in \values$ the pair of the values $v_1$ and $v_2$ and by $(T_1, T_2) \in \types$ the pair of types $T_1$ and $T_2$. We assume \hastype{(v_1, v_2)}{(T_1, T_2)} if and only if \hastype{v_1}{T_1} and \hastype{v_2}{T_2}.
We denote by $\fun{T_1}{T_2}$ the set of total functions from values of type $T_1$ to values of type $T_2$.

We use $\nil$ for the empty sequence, $\app{xs_1}{xs_2}$ for concatenation,
and $\cons{x}{xs}$ for the prepending of $x$ to $xs$.

\subsection{Tokens and Kinds}

We consider a single type $\tokentype \in \types$ to be the type of tokens.
The values $v \in \values$ such that $\hastype{v}{\tokentype}$ are called \emph{tokens}.
We will generally use the lower case letter $t$ to denote such tokens.
The task of parsing consists in turning a sequence of tokens into a value, or to fail when the sequence is invalid.

Each token $t$ is assigned to a single \emph{kind} $\getkind{t}$.
Token kinds represent (potentially infinite) groups of tokens.
We denote by $\kinds$ the set of all kinds, all assumed to be non-empty.
There can be infinitely many different tokens, but
only a finite, relatively small, number of kinds.

As an example, the strings \texttt{"hello world"}, \texttt{"foo"} and \texttt{"bar"} could be considered tokens,
and \texttt{string} would be their token kind. The numbers $3, 17, 42$ could be considered tokens,
while \texttt{number} would be their associated kind.

Token kinds are meant to abstract away details that are irrelevant for recognition.
During parsing, the kinds alone are sufficient to decide whether or not a sequence of tokens is recognised.
However, and importantly, the resulting value built by the parser might depend on the actual tokens.

\subsection{Syntaxes}

For every type $T \in \types$, we define the set $\syntax{T}$ of syntaxes that associates token sequences with values of type $T$.
Those sets are inductively defined by the rules in~\Cref{fig:syntaxes}.

\begin{figure}[H]
\begin{align*}
\inference{k \in \kinds}{\elem{k} \in \syntax{\tokentype}}
&
\hfill
&
\inference{T \in \types}{\fail \in \syntax{T}}
&
\hfill
&
\inference{\hastype{v}{T}}{\eps{v} \in \syntax{T}}
\end{align*}

\begin{align*}
\inference{s_1 \in \syntax{T} & s_2 \in \syntax{T}}{\disj{s_1}{s_2} \in \syntax{T}}
&
\hfill
&
\inference{s_1 \in \syntax{T_1} & s_2 \in \syntax{T_2}}{\seq{s_1}{s_2} \in \syntax{(T_1, T_2)}}
\end{align*}

\begin{align*}
\inference{s \in \syntax{T_1} & f \in \fun{T_1}{T_2}}{\map{f}{s} \in \syntax{T_2}}
&
\hfill
&
\inference{x \in \vars{T}}{\var{x} \in \syntax{T}}
\end{align*}
\caption{Definition of syntaxes.}
\label{fig:syntaxes}
\end{figure}

The construct $\elem{k}$, $\fail$, and $\eps{v}$ form the basic syntaxes.
Intuitively, $\elem{k}$ represents a single token of kind $k$, $\fail$ represents failure, and $\eps{v}$ represents the empty string.
The value $v$ tagged to $\eps{v}$ represents the value associated to the empty string by the syntax.
This value is important as syntaxes are meant not only to describe a language, but also the value associated with each recognised sequence of tokens.

The constructs $\disj{s_1}{s_2}$ and $\seq{s_1}{s_2}$ respectively represent disjunction and sequencing.
The construct $\map{f}{s}$ represents the application of the function $f$ on values produced by $s$.
Finally, the construct $\var{x}$ represents a reference to a syntax defined in a global environment.
The variables and the environment allow for mutually recursive syntaxes.
We consider, for every type $T$, a finite set of identifiers \vars{T}.
The global environment is a finite mapping, that associates
to each identifier $x \in \vars{T}$ a syntax $\getdef{x} \in \syntax{T}$.

\begin{remark}
We use a global environment instead of the equivalent $\mu$-combinator found in other approaches~\cite{leiss1991towards,Krishnaswami:2019:TAA:3314221.3314625} as it is closer to the representation we used in the implementation.
\end{remark}

\subsection{Semantics of Syntaxes}

Syntaxes associate token sequences with values. The inductive predicate $\matches{s}{ts}{v}$ indicates that the syntax $s$ associates the token sequence $ts$ with the value $v$. The inductive predicate is defined by the rules in~\Cref{fig:semantics}.

\input{fig-semantics.tex}

\begin{theorem}
For any type $T \in \types$, syntax $s \in \syntax{T}$, token sequence $ts$ and value $v \in \values$, if $\matches{s}{ts}{v}$ then $\hastype{v}{T}$.
\end{theorem}

\begin{remark}
  We do not present proofs of theorems in this paper and refer instead the reader to our
  proofs in Coq, discussed in~\Cref{sec:coq}.
  Given the order of theorems we present,
  most proofs follow relatively straightforwardly by induction, with main insight being
  the choice of induction variable and schema.
\end{remark}

\subsection{Example}\label{sec:formalismexample}

\begin{figure}
\centering
\input{exampletree.tex}
\caption{Tree representation of the syntax of $x$.}
\label{fig:exampletree}
\end{figure}
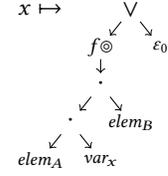

As a simple example of our theoretical framework, we describe a syntax for the mapping $L = \{ \texttt{a}^n \texttt{b}^n\mapsto n\ |\ n \in \mathcal{N} \}$, which assigns to sequences of $n$ $\texttt{a}$'s followed by $n$ $\texttt{b}$'s the integer value $n$.
The tokens we will consider are $\texttt{a}$ and $\texttt{b}$, while their respective kinds are \textit{A} and \textit{B}.
To describe a syntax for this language, we consider the singleton environment that maps the identifier $x$ to (see \Cref{fig:exampletree}):
\[
\begin{gathered}
\disj{\map{f}{(\seq{(\seq{\elem{A}}{\var{x}})}{\elem{B}}})}{\eps{0}}\\
\text{where } f(((t_1, n), t_2)) = n + 1
\end{gathered}
\]

Intuitively, the syntax recognises sequences of the form:
\begin{itemize}
\item a token of kind \textit{A}, followed another instance of the variable $x$, followed by a token of kind \textit{B}, with $f$ applied on the produced value, or
\item the empty sequence of token, with value $0$.
\end{itemize}

In this environment, the mapping $L$ is simply described by the syntax $\var{x}$.
We can derive the following statements about the semantics of the syntax $\var{x}$:
\[
\begin{aligned}
&\matches{\var{x}}{\nil}{0}\hspace{0.5cm}
&\matches{\var{x}}{\sequence{\texttt{a}, \texttt{a}, \texttt{b}, \texttt{b}}}{2}
\end{aligned}
\]
While the following statements are \emph{not} derivable:
\[
\begin{aligned}
&\matches{\var{x}}{\sequence{\texttt{a}, \texttt{a}, \texttt{b}, \texttt{b}}}{17}&
&\matches{\var{x}}{\sequence{\texttt{a}, \texttt{b}, \texttt{a}, \texttt{b}}}{2}&
\\
&\matches{\var{x}}{\sequence{\texttt{a}, \texttt{a}, \texttt{b}}}{2}&
&\matches{\var{x}}{\sequence{\texttt{a}, \texttt{a}, \texttt{b}, \texttt{b}, \texttt{b}}}{2}&
\end{aligned}
\]

The task of parsing is to find out, given a syntax $s$ and a token sequence $ts$, if there exists a $v$ such that $\matches{s}{ts}{v}$, and if so, to return such a $v$.
In the next section, we will introduce properties that characterise a class of syntaxes on which parsing is solvable in worst-case linear time complexity (LL(1) syntaxes).
Afterwards, we will devise a parsing algorithm for such LL(1) syntaxes based on the concept of derivatives.
Finally, we will present an optimisation based on the zipper data structure to obtain worst-case linear time complexity for LL(1) parsing with derivatives.

%% file: fig-semantics.tex
\begin{figure}[H]
\begin{align*}
\inference{k = \getkind{t}}{\matches{\elem{k}}{\singleton{t}}{t}}
&
\hfill
&
\inference{}{\matches{\eps{v}}{\nil}{v}}
\end{align*}
\vspace{0.12cm}
\begin{align*}
\inference{\matches{s_1}{ts}{v}}{\matches{\disj{s_1}{s_2}}{ts}{v}}
&
\hfill
&
\inference{\matches{s_2}{ts}{v}}{\matches{\disj{s_1}{s_2}}{ts}{v}} 
\end{align*}
\vspace{0.12cm}
\[
\inference{\matches{s_1}{ts_1}{v_1} & \matches{s_2}{ts_2}{v_2}}{\matches{\seq{s_1}{s_2}}{\app{ts_1}{ts_2}}{(v_1, v_2)}} 
\]
\vspace{0.12cm}
\[
\inference{\matches{s}{ts}{v}}{\matches{\map{f}{s}}{ts}{f(v)}}
\]
\vspace{0.12cm}
\[
\inference{s = \getdef{x} & \matches{s}{ts}{v}}{\matches{\var{x}}{ts}{v}}
\]
\caption{Semantics of syntaxes.}
\label{fig:semantics}
\end{figure}

%% file: exampletree.tex
\tikzset{snode/.style={minimum size=0.3cm, align=center}}
\tikzset{lnode/.style={minimum size=0.3cm, align=center}}
\tikzset{>=to}

\begin{tikzpicture}[xscale=0.4, yscale=0.5]
\node (x) at (0, 5) {\small$x \mapsto$};
\node[snode, label=center:$\vee$] (dis1) at (3, 5) {};
\node[snode, label=center:\scriptsize$\eps{0}$] (eps1) at (4, 4) {};
\draw[->] (dis1) -- (eps1);
\node[snode, label=center:\scriptsize$f \circledcirc$] (app1) at (2, 4) {};
\draw[->] (dis1) -- (app1);
\node[snode, label=center:$\cdot$] (seq1) at (2, 3) {};
\draw[->] (app1) -- (seq1);
\node[snode, label=center:$\cdot$] (seq2) at (1, 2) {};
\draw[->] (seq1) -- (seq2);
\node[snode, label=center:\scriptsize$\elem{B}$] (elemB1) at (3, 2) {};
\draw[->] (seq1) -- (elemB1);
\node[snode, label=center:\scriptsize$\elem{A}$] (elemA1) at (0, 1) {};
\draw[->] (seq2) -- (elemA1);
\node[snode, label=center:\scriptsize$\var{x}$] (var1) at (2, 1) {};
\draw[->] (seq2) -- (var1);
\end{tikzpicture}

%% file: properties.tex

\section{Properties of Syntaxes}

This section defines several computable properties of syntaxes which we use
for LL(1) checking and parsing.

\subsection{Productivity}

A syntax is said to be \emph{productive} (see \Cref{fig:prod}) if it associates at least one sequence of tokens with a value.

\begin{theorem}
For any syntax $s$:
\[
\prod{s} \iff \exists ts, v.\ \matches{s}{ts}{v}
\]
\end{theorem}

Not all syntaxes are productive. For instance $\fail$, $\seq{\fail}{\elem{k}}$, $\seq{\elem{k}}{\fail}$ are all non-productive. Non-productive syntaxes can also occur due to non-well-founded recursion, as in the syntax $\var{x}$ with an environment mapping $x$ to $\seq{\elem{k}}{\var{x}}$. 

\subsection{Nullability}

A syntax $s \in \syntax{T}$ is said to be \emph{nullable with value $v$} (see \Cref{fig:nullable}), if it associates to the empty sequence of tokens the value $v$ of type $T$.
We will say $s$ is \emph{nullable} when we do not need to refer to the value that $s$ is nullable with.
We will use the function $\nullfun{\_}$ to return an arbitrary nullable value from a syntax, if such a value exists, or $\none$ otherwise.

\begin{theorem}
For any syntax $s$ and value $v$:
\[
\nullwith{s}{v} \iff \matches{s}{\nil}{v}
\]
\end{theorem}

\subsection{First Set}

The \emph{first set} of a syntax $s$ (see \Cref{fig:first}) is the set containing the kinds of all tokens at the start of at least one sequence associated with some value by $s$. 

\begin{theorem}
The first set of a syntax $s$ equals the set
\[
\{\ k\ |\ \exists t, ts, v.\ \getkind{t} = k \wedge \matches{s}{\cons{t}{ts}}{v}\ \}
\]
\end{theorem}

\subsection{Should-Not-Follow Set}

The concept of a \emph{should-not-follow set} is directly connected to the concept of LL(1) conflicts that we will later introduce.
Intuitively, the should-not-follow set of a syntax is the set of kinds that would introduce an ambiguity if the first set of any syntax directly following that  syntax was to contain that kind.
The concept of should-not-follow set is used as an alternative to the concept of \texttt{FOLLOW} set generally used in the context of LL(1) parsing.
While the \texttt{FOLLOW} set is a global property of a grammar, the should-not-follow set enjoys a local nature.
Contrary to its \texttt{FOLLOW} set, the should-not-follow set of a syntax is a property of the syntax alone, not of its surrounding context.
We define the should-not-follow set inductively in~\Cref{fig:snf}.
Our definition differs from the one used by~\citet{Krishnaswami:2019:TAA:3314221.3314625} and introduced in earlier works~\cite{johnstone1998generalised, bruggemann1992deterministic}.
While we introduce elements to the set in the case of disjunctions, they do so in the case of sequences.
Our definition seems more appropriate: the previous work introduced additional restrictions on syntaxes,
disallowing nullable expressions on left part of sequence, which is \emph{not} needed in our approach (nor in conventional LL(1) definition for context-free grammars
\cite[Theorem 5.3, Page 343]{DBLP:books/lib/AhoU72}).

\begin{theorem} \label{thm:snfollow-sound}
For any syntax $s$ and kind $k$, if $k$ is part of the should-not-follow set of $s$, then there exist a token $t$ of kind $k$ and (possibly empty) sequences of token $ts_1$ and $ts_2$ such that:
\begin{align*}
&\matches{s}{ts_1}{v_1}\hspace{1em}\wedge \hspace{1em}
\matches{s}{\app{ts_1}{(\cons{t}{ts_2}})}{v_2}&
\end{align*}
\end{theorem}

\input{fig-properties}

\subsection{LL(1) Conflicts}

Finally, we introduce in~\Cref{fig:conflicts} the notion of LL(1) conflicts.
When a syntax has LL(1) conflicts, a choice between two alternatives can arise during parsing which can not be resolved given a single token of lookahead.
Informally, LL(1) conflicts arise in three cases:
1) Both branches of a disjunction are nullable, which means that two potentially different values are associated with the empty string by the disjunction. 
2) Branches of a disjunction have non-disjoint first sets, so both branches can accept a sequence starting with the same token.
Given a single token of lookahead, a parser thus cannot decide which branch to choose.
3) The should-not-follow set of the left-hand side of a sequence and the
first set of the right-hand side of that sequence both contain the same token
kind, $k$. This means that there is a point in the left-hand side (after reading
$ts_1$ from \Cref{thm:snfollow-sound}) where reading a
token of kind $k$ will make it impossible to decide whether we should stay in
the left-hand side (and then read $ts_2$), or start parsing in the right-hand
side.
\begin{definition}
A syntax is LL(1) iff it has no LL(1) conflicts.
\end{definition}

\begin{theorem}\label{thm:nonambiguous}
  [LL(1) syntaxes are non-ambiguous] For all LL(1) syntaxes $s$, token sequences $ts$ and values $v_1$ and $v_2$:
  \begin{gather*}
  \matches{s}{ts}{v_1}\hspace{0.3em} \wedge \hspace{0.3em}\matches{s}{ts}{v_2} \implies v_1 = v_2
  \end{gather*}
\end{theorem}

As a direct consequence of \Cref{thm:nonambiguous}, we have that there can be at most a single nullable value associated to any LL(1) syntax, and thus that $\nullfun{\_}$ is completely deterministic for LL(1) syntaxes.

\begin{theorem}
Should-not-follow set of an LL(1) syntax $s$ is
\begin{align*}
\{\ k\ |\ \exists t, ts_1, ts_2, v_1, v_2.\ &\getkind{t} = k\ \wedge\\
&\matches{s}{ts_1}{v_1}\ \wedge\\
&\matches{s}{\app{ts_1}{(\cons{t}{ts_2}})}{v_2}\ \}
\end{align*}
\end{theorem}

\subsection{Left-Recursivity}

We say an identifier $x$ is \emph{left-recursive} if $\var{x}$ can be visited within $\getdef{x}$ without consuming input tokens.
Formally, an identifier $x$ is left-recursive if $x \in \visitable{\getdef{x}}$, where rules for \emph{visitability} are given by~\Cref{fig:visitable}.

\begin{figure}
\[
\inference{}{x \in \visitable{\var{x}}}
\]
\vspace{0.2cm}
\begin{align*}
\inference{x \in \visitable{s_1}}{x \in \visitable{\disj{s_1}{s_2}}}
&
\hfill
&
\inference{x \in \visitable{s_2}}{x \in \visitable{\disj{s_1}{s_2}}}
\end{align*}
\vspace{0.2cm}
\[
\inference{x \in \visitable{s_1}}{x \in \visitable{\seq{s_1}{s_2}}}
\]
\vspace{0.2cm}
\[
\inference{\nullwith{s_1}{v} & x \in \visitable{s_2}}{x \in \visitable{\seq{s_1}{s_2}}}
\]
\vspace{0.2cm}
\[
\inference{x \in \visitable{s}}{x \in \visitable{\map{f}{s}}}
\]
\vspace{0.2cm}
\[
\inference{s = \getdef{y} & x \in \visitable{s}}{x \in \visitable{\var{y}}}
\]
\caption{Rules for inclusion in the visitable set.}
\label{fig:visitable}
\end{figure}

\begin{theorem}\label{thm:leftrec}
For any left-recursive identifier $x$, when $\getdef{x}$ is productive, the syntax $\getdef{x}$ is not LL(1). 
\end{theorem}

\subsection{Computing with Propagator Networks}\label{sec:networks}

The definitions we introduced in this section are based on inductive rules.
Due to the cyclic nature of syntaxes arising from the variables and global environment,
those definitions do not immediately give rise to recursive procedures.
The usual solution to this problem is to iteratively apply rules until a fix-point is reached.
We instead propose using \emph{propagator networks}~\cite{steele1980definition, radul2009propagation} as a principled and efficient way to compute the properties defined in the present section.
The idea is to build a network of \emph{cells}, one for each node in the syntax.
For each identifier $x$, the $\var{x}$ nodes share the same cell.
Each cell has a mutable state which holds information about the properties of the corresponding syntax node.
We maintain a list of cells that need to be updated.
Information is propagated through the network by
updating the content of such cells according to the inductive rules presented in~\Cref{fig:defs}.
Using this approach, we found that properties can be computed for a syntax and
all its inner nodes in worst-case time linear in the size of the syntax, which is not direct from the conventional fix-point definitions.
The constant number of kinds also factors in the cost of computations of first and should-not-follow sets.
We have proven the correctness of the approach in Coq, as further discussed in~\Cref{sec:coq}.

%% file: fig-properties.tex
\begin{figure*}[tbp]
\begin{subfigure}{0.45\textwidth}
\begin{align*}
\inference{}{\prod{\eps{v}}}
&
\hfill
&
\inference{}{\prod{\elem{k}}}
\end{align*}
\vspace{0.2cm}
\begin{align*}
\inference{\prod{s_1}}{\prod{\disj{s_1}{s_2}}}
&
\hfill
&
\inference{\prod{s_2}}{\prod{\disj{s_1}{s_2}}}
\end{align*}
\vspace{0.2cm}
\[
\inference{\prod{s_1} & \prod{s_2}}{\prod{\seq{s_1}{s_2}}} 
\]
\vspace{0.2cm}
\[
\inference{\prod{s}}{\prod{\map{f}{s}}}
\]
\vspace{0.2cm}
\[
\inference{s = \getdef{x} & \prod{s}}{\prod{\var{x}}}
\]
\caption{Rules for productivity.}
\label{fig:prod}
\end{subfigure}
\hfill
\begin{subfigure}{0.45\textwidth}
\[
\inference{}{\nullwith{\eps{v}}{v}}
\]
\vspace{0.2cm}
\begin{align*}
\inference{\nullwith{s_1}{v}}{\nullwith{\disj{s_1}{s_2}}{v}}
&
\hfill
&
\inference{\nullwith{s_2}{v}}{\nullwith{\disj{s_1}{s_2}}{v}}
\end{align*}
\vspace{0.2cm}
\[
\inference{\nullwith{s_1}{v_1} & \nullwith{s_2}{v_2}}{\nullwith{\seq{s_1}{s_2}}{(v_1, v_2)}} 
\]
\vspace{0.2cm}
\[
\inference{\nullwith{s}{v}}{\nullwith{\map{f}{s}}{f(v)}}
\]
\vspace{0.2cm}
\[
\inference{s = \getdef{x} & \nullwith{s}{v}}{\nullwith{\var{x}}{v}}
\]
\caption{Rules for nullability.}
\label{fig:nullable}
\end{subfigure}

\vspace{0.2cm}

\begin{subfigure}{0.45\textwidth}
\[
\inference{}{k \in \first{\elem{k}}}
\]
\vspace{0.2cm}
\begin{align*}
\inference{k \in \first{s_1}}{k \in \first{\disj{s_1}{s_2}}}
&
\hfill
&
\inference{k \in \first{s_2}}{k \in \first{\disj{s_1}{s_2}}}
\end{align*}
\vspace{0.2cm}
\[
\inference{k \in \first{s_1} & \prod{s_2}}{k \in \first{\seq{s_1}{s_2}}}
\]
\vspace{0.2cm}
\[
\inference{\nullwith{s_1}{v} & k \in \first{s_2}}{k \in \first{\seq{s_1}{s_2}}}
\]
\vspace{0.2cm}
\[
\inference{k \in \first{s}}{k \in \first{\map{f}{s}}}
\]
\vspace{0.2cm}
\[
\inference{s = \getdef{x} & k \in \first{s}}{k \in \first{\var{x}}}
\]
\caption{Rules for inclusion in the first set.}
\label{fig:first}
\end{subfigure}
\hfill
\begin{subfigure}{0.45\textwidth}
\begin{align*}
\inference{k \in \snf{s_1}}{k \in \snf{\disj{s_1}{s_2}}}
&
\hfill
&
\inference{k \in \snf{s_2}}{k \in \snf{\disj{s_1}{s_2}}}
\end{align*}
\vspace{0.2cm}
\[
\inference{k \in \first{s_1} & \nullwith{s_2}{v}}{k \in \snf{\disj{s_1}{s_2}}}
\]
\vspace{0.2cm}
\[
\inference{\nullwith{s_1}{v} & k \in \first{s_2}}{k \in \snf{\disj{s_1}{s_2}}}
\]
\vspace{0.2cm}
\[
\inference{k \in \snf{s_1} & \nullwith{s_2}{v}}{k \in \snf{\seq{s_1}{s_2}}}
\]
\vspace{0.2cm}
\[
\inference{\prod{s_1} & k \in \snf{s_2}}{k \in \snf{\seq{s_1}{s_2}}}
\]
\vspace{0.2cm}
\[
\inference{k \in \snf{s}}{k \in \snf{\map{f}{s}}}
\]
\vspace{0.2cm}
\[
\inference{s = \getdef{x} & k \in \snf{s}}{k \in \snf{\var{x}}}
\]
\caption{Rules for inclusion in the should-not-follow set.}
\label{fig:snf}
\end{subfigure}

\vspace{0.2cm}

\begin{subfigure}{\textwidth}
\begin{align*}
\inference{\nullwith{s_1}{v_1} & \nullwith{s_2}{v_2}}{\conflict{\disj{s_1}{s_2}}}
&
\hfill
&
\inference{k \in \first{s_1} & k \in \first{s_2}}{\conflict{\disj{s_1}{s_2}}}
&
\hfill
&
\inference{k \in \snf{s_1} & k \in \first{s_2}}{\conflict{\seq{s_1}{s_2}}}
\end{align*}
\vspace{0.2cm}
\begin{align*}
\inference{\conflict{s_1}}{\conflict{\disj{s_1}{s_2}}}
&
\hfill
&
\inference{\conflict{s_2}}{\conflict{\disj{s_1}{s_2}}}
&
\hfill
&
\inference{\conflict{s_1}}{\conflict{\seq{s_1}{s_2}}}
&
\hfill
&
\inference{\conflict{s_2}}{\conflict{\seq{s_1}{s_2}}}
\end{align*}
\vspace{0.2cm}
\begin{align*}
\inference{\conflict{s}}{\conflict{\map{f}{s}}}
&
\hfill
&
\inference{s = \getdef{x} & \conflict{s}}{\conflict{\var{x}}}
\end{align*}
\caption{Rules for existence of LL(1) conflicts.}
\label{fig:conflicts}
\end{subfigure}
\caption{Inductive definitions of properties on syntaxes.}
\label{fig:defs}
\end{figure*}

%% file: derivative.tex

\section{Simple LL(1) Parsing with Derivatives}\label{sec:simplederive}

In this section, we introduce the concept of derivatives of LL(1) syntaxes and show how that concept leads to a simple parsing algorithm.
Later, in~\Cref{{sec:zippy}}, we discuss inefficiencies of that algorithm and propose a crucial optimisation which makes the algorithm run in linear time.

\begin{remark}
Theorems of this section are omitted from the Coq formalisation discussed in~\Cref{sec:coq}, as they are not relevant to the correctness of the parsing algorithm presented in~\Cref{sec:zippy}.
\end{remark}

\subsection{Derivatives of LL(1) Syntaxes}\label{sec:ll1derive}

The \emph{derivative} of a syntax $s$ with respect to a token $t$ is a new syntax $\derive{t}{s}$ which associates for every sequence $ts$ the value $v$ if and only if $s$ associates $\cons{t}{ts}$ with $v$. The derivative of a syntax with respect to a token represents the state of the syntax after seeing the token $t$. Instead of defining the derivative for the general case, we only define it for LL(1) syntaxes $s$ and tokens $t$
such that $\getkind{t} \in \first{s}$:

\begin{align*}
\derive{t}{\elem{k}} &:= \eps{t}\\
\derive{t}{\disj{s_1}{s_2}} &:= \begin{cases}
\derive{t}{s_1} & \text{if $\getkind{t} \in \first{s_1}$} \\
\derive{t}{s_2} & \text{otherwise}
\end{cases}
\\
\derive{t}{\seq{s_1}{s_2}} &:= \begin{cases}
\seq{\eps{v}}{\derive{t}{s_2}} & \text{if $\nullfun{s_1} = \some{v}$}\\
& \text{and $\getkind{t} \in \first{s_2}$} \\
\seq{\derive{t}{s_1}}{s_2} & \text{otherwise}
\end{cases}
\\
\derive{t}{\map{f}{s}} &:= \map{f}{\derive{t}{s}}
\\
\derive{t}{\var{x}} &:= \derive{t}{\getdef{x}}
\end{align*}

\vspace{0.1cm}

The invariant that the token kind must be part of the first set reduces the number of cases to consider.
In addition, the restriction to LL(1) syntaxes allows for drastic simplifications.
Compared to the original definition of derivatives of context-free expressions by~\citet{might2011parsing},
our definition only performs recursive calls on at most one child syntax.
The choice of which child to recursively derive is informed by first sets.
Thanks to~\Cref{thm:leftrec}, variables that are derived are not left-recursive, so the recursion is well-founded.

\begin{theorem}
The syntax $\derive{t}{s}$ is well-defined for any LL(1) syntax $s$ and token $t$ of kind $k \in \first{s}$.
\end{theorem}

\begin{theorem}[Progress]
  For any LL(1) syntax $s$, token $t$ of kind $k \in \first{s}$, token sequence $ts$ and value $v$ we have that
$s$ associates the token sequence $\cons{t}{ts}$ with the value $v$   iff
  $\derive{t}{s}$ associates the token sequence $ts$ with the same value $v$:
\begin{gather*}
\forall s, t.\ 
\neg\conflict{s}
\wedge
\getkind{t} \in \first{s}
\implies\\
\forall ts, v.\ 
\matches{s}{\cons{t}{ts}}{v}
\iff
\matches{\derive{t}{s}}{ts}{v}
\end{gather*}
\end{theorem}

\begin{theorem}[Preservation]
For any LL(1) syntax $s$ and token $t$ of kind $k \in \first{s}$, the syntax $\derive{t}{s}$ is LL(1).
\\
\begin{gather*}
\forall s, t.\ 
\neg\conflict{s}
\wedge
\getkind{t} \in \first{s}
\implies\\
\neg\conflict{\derive{t}{s}}
\end{gather*}
\end{theorem}

\subsection{Simple LL(1) Parsing with Derivatives} \label{sec:simplicito}

The derivation operation naturally leads to a parsing algorithm for LL(1) syntaxes:

\begin{minipage}{\linewidth}
\begin{lstlisting}
!*$\parse{s}{\nil}$\hspace{1.25em}*! := !*$\nullfun{s}$*!
!*$\parse{s}{\cons{t}{ts}}$*! := if !*$\getkind{t} \in \first{s}$*!
!*\hspace{7.8em}*!then !*$\parse{\derive{t}{s}}{ts}$*! else !*$\none$*!
\end{lstlisting}
\end{minipage}

\begin{theorem}[Correctness]
For any LL(1) syntax $s$, token sequence $ts$ and value $v$:
\[
\parse{s}{ts} = \some{v}
\iff
\matches{s}{ts}{v}
\]
\end{theorem}

%% file: zippy.tex

\section{Zippy LL(1) Parsing with Derivatives}
\label{sec:zippy}

In this section, we demonstrate that the simple parsing with derivatives for LL(1) syntaxes of \Cref{sec:simplicito} can have bad performance.
To alleviate this problem, we introduce the concept of \emph{focused syntaxes}, which combine a syntax and a \emph{context}.
We show that, using such ``zipper'' data structure~\cite{huet1997zipper},
LL(1) parsing with derivatives takes linear time.

\subsection{Inefficiency of Simple Parsing with Derivatives}\label{sec:inefficient}

While correct, parsing with derivatives as shown in the previous section is inefficient in practice.
As we will show, the derivative of a syntax can grow larger than the original syntax.
Partially created values, as well as continuation points, will tend to accumulate in the top layers of the syntax.
With time, calls to the derive procedure will take longer and longer as they will need to traverse deeper and deeper syntaxes.
It can be shown that the parsing algorithm that we have described in the previous section takes time quadratic in the input size because of that phenomenon,
whereas the typical push-down automaton-based parsing algorithm for LL(1) grammars only takes linear time
\cite{DBLP:books/lib/AhoU72}.
Furthermore, simple parsing with derivatives can lead to stack overflows because derivation as defined above is not tail-recursive.

\paragraph*{Example}

As a simple example to expose the problematic behaviour of the algorithm, let us come back to the example mapping $L = \{ \texttt{a}^n \texttt{b}^n\mapsto n\ |\ n \in \mathcal{N} \}$ introduced in~\Cref{sec:formalismexample}.
In this example, the environment consists of a single entry:
\[
\begin{gathered}
x \mapsto \disj{\map{f}{(\seq{(\seq{\elem{A}}{\var{x}})}{\elem{B}}})}{\eps{0}}\\
\text{where } f(((t_1, n), t_2)) = n + 1
\end{gathered}
\]
The syntax that describes $L$ is $\var{x}$.
The syntax is LL(1).

To showcase the problematic behaviour of the algorithm, define the following sequence of syntaxes:
\begin{gather*}
s_0 := \var{x}\hspace{1.5em}
s_{i + 1} := \derive{\texttt{a}}{s_i}
\end{gather*}
The first element of the sequence $s$ is the original syntax $\var{x}$, while subsequent elements are derivatives of the previous syntax with respect to $\texttt{a}$.
This sequence models the state of the parsing with derivatives algorithm after encountering longer and longer strings of $\texttt{a}$'s.
\Cref{fig:syntaxtrees}~shows the trees corresponding to $s_1$, $s_2$ and $s_3$.

\input{trees.tex}

We immediately observe that $s_i$ grows larger and larger as $i$ grows.
Each time a new $\texttt{a}$ is encountered, additional nodes are added at the \emph{bottom} of the previous syntax in place of the $\var{x}$ sub-syntax.
Indeed, the derivation process works its way down the syntax, unfolding variables as needed, until an $\elem{A}$ node is found. At that point, the located $\elem{A}$ node is replaced by an $\eps{\texttt{a}}$ node.
As we can observe, the $\elem{A}$ node (hidden within $\var{x}$) is located deeper and deeper within each $s_i$, making the derivation process  longer and longer.
In addition to that, as syntaxes are immutable structures, replacing the located $\elem{A}$ node forces each node from the root to the leaf node $\elem{A}$ to be copied.
Therefore, we can easily observe that computing the derivative of $s_i$ takes time linear in $i$.
What this means is that, in this particular case, the parsing algorithm that we have discussed in the previous section would require time \emph{quadratic} in the input size, which is not desirable for LL(1) languages.

\begin{figure}
\centering
\input{focusedtree.tex}
\caption{Representation of the syntax $s_3$ with a focus on the bottommost $\eps{\texttt{a}}$ node.}
\label{fig:focusedtree}
\end{figure}

To tackle this phenomenon, we introduce \emph{focused syntaxes}.
The idea is very simple:
Instead of having pointers always flowing away from the root down to the leafs,
we will use a different structure in which pointers flow away from an inner syntax node called the \emph{focal point}.
To represent the syntax nodes on the path from the root of the tree to the focal point, we will introduce the concept of \emph{layers}.
\Cref{fig:focusedtree} shows the focused syntax corresponding to $s_3$, with the bottommost $\eps{\texttt{a}}$ as the focal point.
In the next section, we will formalise this concept and show how to adapt the LL(1) parsing with derivatives algorithm to this zipper-inspired structure.

\subsection{Focused Syntaxes}

A focused syntax is simply a syntax with a focus on one of its nodes, in the spirit of zippers~\cite{huet1997zipper}.
We define a \emph{focused syntax} as a pair of a syntax $s$ and a stack of \emph{layers} $c$.
Given a focused syntax $(s, c)$, we call $s$ the \emph{focal point} and $c$ the \emph{context}.
Layer of the context are of three different forms:
\begin{itemize}
\item
$\apply{f}$, which indicate that a function $f$ is to be applied on the parsed value.
\item
$\prepend{v}$, which indicate that a value $v$ must be prepended to the parsed value.
\item
$\followby{s}$, which indicate that the syntax $s$ follows in sequence.
\end{itemize}
These correspond to the syntax nodes that can be created by the LL(1) derivation procedure shown in~\Cref{sec:simplederive}:
\begin{itemize}
\item
$\apply{f}$ corresponds to $\map{f}{\_}$,
\item
$\prepend{v}$ corresponds to $\seq{\eps{v}}{\_}$,
\item
$\followby{s}$ corresponds to $\seq{\_}{s}$.
\end{itemize}
Note that there are no layers for disjunctions, as such layers are not introduced by the LL(1) derivation procedure.
In addition, note that there can not be any actual syntax to the left of the focal point.
Indeed, our limited set of layers will ensure the focal point is always \emph{to the left} of the syntax.

Layers are parameterised by two types, the \emph{above} type and the \emph{below} type. We denote by $\layers{T_1}{T_2}$ the set of all layers with above type $T_1$ and below type $T_2$ (see \Cref{fig:layers}).

\begin{figure}
\begin{align*}
\inference{f \in \fun{T_1}{T_2}}{\apply{f} \in \layers{T_1}{T_2}}
&
\hfill
&
\inference{\hastype{v}{T_1}}{\prepend{v} \in \layers{T_2}{(T_1, T_2)}}
\end{align*}
\vspace{0.18cm}
\[
\inference{s \in \syntax{T_2}}{\followby{s} \in \layers{T_1}{(T_1, T_2)}}
\]
\caption{Definition of layers.}
\label{fig:layers}
\end{figure}

The context is a stack of type-aligned layers.
For any two consecutive layers in the stack, the below type of the first layer must match the above type of the second layer.
For any types $T_1$ and $T_2$, we denote by $\context{T_1}{T_2}$ the set of all type-aligned contexts where $T_1$ is the above type of the first layer and $T_2$ is the below type of the last layer. 
For all types $T$ we also include the empty stack $\nil$ in $\context{T}{T}$.

A focused syntax in $\focustype{T}$ is a pair of a syntax $s \in \syntax{T'}$ and a context $c \in \context{T'}{T}$ for some type $T'$.
We define the function \texttt{focus} to focus the root node of a syntax:

\begin{minipage}{\linewidth}
\begin{lstlisting}
!*\focus{s}*! := !*$(s, \nil)$*!
\end{lstlisting}
\end{minipage}

Conversely, the \texttt{unfocus} function on a focused syntax $(s, c)$ in $\focustype{T}$ unfocuses $(s,c)$ by
applying the layers in the context until the context is empty:

\begin{minipage}{\linewidth}
\begin{lstlisting}
!*\unfocus{(s, c)}*! := match c with
  | !*$\nil$*! !*\hspace{6.5em}*! !*$\to$*! !*$s$*!
  | !*$\cons{\apply{f}}{c'}$*! !*\hspace{1.8em}*! !*$\to$*! !*$ \unfocus{(\map{f}{s}, c')}$*!
  | !*$\cons{\prepend{v}}{c'}$*! !*\hspace{0.85em}*! !*$\to$*! !*$\unfocus{(\seq{\eps{v}}{s}, c')}$*!
  | !*$\cons{\followby{s'}}{c'}$*! !*$\to$*! !*$\unfocus{(\seq{s}{s'}, c')}$*!
\end{lstlisting}
\end{minipage}

\begin{definition}
A focused syntax is LL(1) if its unfocused counterpart is LL(1).
\end{definition}

\subsection{Operations on Focused Syntaxes}

In this section, we introduce several operations on focused syntaxes towards designing an efficient parsing procedure.
The first operation we define on focused syntaxes is \texttt{plug}.
The goal of this operation is to obtain a new focused syntax when the focal point reduces down to a value.
This happens for instance when the focal point is an $\eps{v}$ syntax.
The function takes as input a value and a context, and returns a new focused syntax.
Layers in the context are applied until either a $\followby{s}$ layer is encountered, or until the context is empty.

\begin{minipage}{\linewidth}
\begin{lstlisting}
!*\plug{v}{c}*! := match c with
  | !*$\nil$*! !*\hspace{6.2em}*! !*$\to$*! !*$(\eps{v}, \nil)$*!
  | !*$\cons{\apply{f}}{c'}$*! !*\hspace{1.5em}*! !*$\to$*! !*$\plug{f(v)}{c'}$*!
  | !*$\cons{\prepend{v'}}{c'}$*! !*\hspace{0.25em}*! !*$\to$*! !*$\plug{(v', v)}{c'}$*!
  | !*$\cons{\followby{s}}{c'}$*! !*$\to$*! !*$(s, \cons{\prepend{v}}{c'})$*!
\end{lstlisting}
\end{minipage}

\begin{theorem}
  The focused syntax obtained by plugging a value $v$ in a context $c$ is equivalent to $(\eps{v}, c)$. Formally:
  $\forall ts, w$,
\[\begin{array}{rl}
\matches{\unfocus{\plug{v}{c}}}{ts}{w}   &\iff\\
 \matches{\unfocus{(\eps{v}, c)}}{ts}{w}
\end{array}\]
\end{theorem}

The next operation we define is $\texttt{locate}$, which
takes as input a token kind and a focused syntax, and returns an optional 
focused syntax.
The goal of the function is to move the focus towards a syntax node that can start with a given token kind, skipping nullable prefixes as needed.

\begin{minipage}{\linewidth}
\begin{lstlisting}
!*$\locate{k}{(s, c)}$*! :=
  if !*$k \in \first{s}$*! then !*$\some{(s, c)}$*!
  else match !*$\nullfun{s}$*! with
    | !*\none*! !*$\to$*! !*\none*!
    | !*$\some{v}$*! !*$\to$*! if !*$c = \nil$*! then !*\none*!
                        else !*$\locate{k}{\plug{v}{c}}$*!
\end{lstlisting}
\end{minipage}

\noindent
In case the current focal point starts with the desired kind the current focused syntax is simply returned.
Otherwise, the focus is to be moved to a consecutive syntax found within the context, at which point the operation is recursively applied.
Note that the operation does not always succeed, and so for two reasons.
First, in order to be able to skip the currently focused node, that node must be nullable.
Second, the context might be empty, and therefore no consecutive syntax exists.

\begin{theorem}
When the \texttt{locate} function returns $\none$, the focused syntax can not possibly start with the desired kind.
\[
\begin{gathered}
\locate{k}{(s, c)} = \none \implies k \not\in \first{\unfocus{(s, c)}}
\end{gathered}
\]
\end{theorem}

\begin{theorem}
For any focused syntax $(s, c)$ and token kind $k$, when $\texttt{locate}$ successfully returns a new focused syntax, the new focal point starts with the given token kind $k$.
\[
\begin{gathered}
\locate{k}{(s, c)} = \some{(s', c')} \implies k \in \first{s'}
\end{gathered}
\]
\end{theorem}

\begin{figure*}
\input{execution.tex}
\caption{Example execution of the zippy LL(1) parsing with derivatives algorithm on the example focused syntax with input tokens $\sequence{\texttt{a}, \texttt{a}, \texttt{b}, \texttt{b}}$. The focused syntax after each call to \texttt{locate} and \texttt{pierce} is shown.}
\label{fig:execution}
\end{figure*}

\begin{theorem}
For any LL(1) focused syntax $(s, c)$, token $t$ and associated kind $k = \getkind{t}$, when $\texttt{locate}$ successfully returns a new focused syntax, then that focused syntax is equivalent for all sequences that start with the token $t$.
\[
\begin{gathered}
\locate{k}{(s, c)} = \some{(s', c')} \implies
\forall ts, v.\\
\begin{array}{rl}
\matches{\unfocus{(s', c')}}{\cons{t}{ts}}{v}   &\iff\\
\matches{\unfocus{(s, c)}}{\cons{t}{ts}}{v}
\end{array}
\end{gathered}
\]
\end{theorem}

The next operation we consider is \texttt{pierce}.
Given a LL(1) syntax $s$ and a token kind $k$ where $k \in \first{s}$, the function returns the \emph{context} around the unique $\elem{k}$ in a left-most position in $s$.
An initial accumulator context is given to the function, and is only built upon by \texttt{pierce}.

\begin{minipage}{\linewidth}
\begin{lstlisting}
!*$\pierce{k}{s}{c}$*! := match s with
  | !*$\elem{k}$*! !*$\to$*! c
  | !*$\disj{s_1}{s_2}$*! !*$\to$*!
    if !*$k \in \first{s_1}$*! then !*$\pierce{k}{s_1}{c}$*!
    else !*$\pierce{k}{s_2}{c}$*!
  | !*$\seq{s_1}{s_2}$*! !*$\to$*!
    match !*$\nullfun{s_1}$*! with
    | !*$\none$*! !*$\to$*! !*$\pierce{k}{s_1}{\cons{\followby{s_2}}{c}}$*!
    | !*$\some{v}$*! !*$\to$*!
      if !*$k \in \first{s_1}$*!
      then !*$\pierce{k}{s_1}{\cons{\followby{s_2}}{c}}$*!
      else !*$\pierce{k}{s_2}{\cons{\prepend{v}}{c}}$*!
  | !*$\map{f}{s'}$*! !*$\to$*! !*$\pierce{k}{s'}{\cons{\apply{f}}{c}}$*!
  | !*$\var{x}$*! !*$\to$*! !*$\pierce{k}{\getdef{x}}{c}$*!
\end{lstlisting}
\end{minipage}

\vspace{0.1cm}

\noindent
The recursive structure of this operation is similar to the one of the derivation operation on LL(1) syntaxes that we have presented in~\Cref{sec:ll1derive}.
The function \texttt{pierce} can be thought of as computing the derivative of a LL(1) syntax, but instead of directly building the resulting syntax, the function returns an equivalent context around the $\elem{k}$ at the base of the recursion.

\begin{theorem}
For any LL(1) focused syntax $(s, c)$ and token~$t$ of kind $k$ where $k \in \first{s}$, the following holds:
\begin{align*}
\forall ts, v.\
&\matches{\unfocus{(\elem{k}, \pierce{k}{s}{c})}}{\cons{t}{ts}}{v} \iff{}\\
&\matches{\unfocus{(s, c)}}{\cons{t}{ts}}{v}
\end{align*}
\end{theorem}

\begin{theorem}\label{thm:derivationtrivial}
Once focused on an $\elem{k}$ node, derivation is trivial.
For any context $c$ and token $t$ of kind $k$:
\begin{align*}
\forall ts, v.\
&\matches{\unfocus{(\elem{k}, c)}}{\cons{t}{ts}}{v} \iff{}\\
&\matches{\unfocus{(\eps{t}, c)}}{ts}{v}
\end{align*}
\end{theorem}

Finally, the function \texttt{derive} brings the various operations we have seen so far together.
The function takes as argument a token $t$ and an LL(1) focused syntax $(s, c)$.
The function returns a new focused syntax $(s', c')$ that corresponds to the derivative of $(s, c)$ with respect to $t$,
or $\none$ if the token is not accepted by the focused syntax.

\begin{minipage}{\linewidth}
\begin{lstlisting}
!*$\fderive{t}{(s, c)}$*! := let !*$k$*! := !*$\getkind{t}$*! in
  match !*$\locate{k}{(s, c)}$*! with
  | !*\none*! !*$\hspace{2.85em}$*! !*$\to$*! !*\none*!
  | !*$\some{(s', c')}$*! !*$\to$*! !*$\some{(\eps{t}, \pierce{k}{s'}{c'})}$*!
\end{lstlisting}
\end{minipage}
\noindent
The operation first invokes \texttt{locate} to move the focus to a point which starts with the desired kind $k$,
then, using \texttt{pierce}, moves the focus down to the left-most $\elem{k}$ within that syntax.
Once focused on that particular $\elem{k}$ node, derivation is trivial, as it suffices to replace the focal point by an $\eps{t}$ node.

\begin{theorem}
The \texttt{derive} operation preserves the LL(1)-ness of the focused syntax.
In other words, for any LL(1) focused syntax $(s, c)$, if its derivation exists, then the resulting focused syntax is also LL(1).
\end{theorem}

\begin{theorem}
When the \texttt{derive} operation returns $\none$ for a token $t$ (of kind $k$) and a focused syntax $(s, c)$ then the corresponding unfocused syntax doesn't start with $k$.
\[
\begin{gathered}
\fderive{t}{(s, c)} = \none \implies
k \not\in \first{\unfocus{(s, c)}}
\end{gathered}
\]
\end{theorem}

\begin{theorem}
For all LL(1) focused syntax $(s, c)$ and token $t$, if the derivation returns a new focused syntax $(s', c')$, then $(s', c')$ is the derivative of $(s, c)$ with respect to $t$.
\[
\begin{gathered}
\fderive{t}{(s, c)} = \some{(s', c')}
\implies
\forall ts, v.\\
\matches{\unfocus{(s', c')}}{ts}{v}
\Leftrightarrow
\matches{\unfocus{(s, c)}}{\cons{t}{ts}}{v}
\end{gathered}
\]

\label{thm:derive-lang}
\end{theorem}

The final piece of the puzzle is the $\texttt{result}$ operation, which returns the value associated with the empty string by the focused syntax.

\begin{minipage}{\linewidth}
\begin{lstlisting}
!*$\fnullable{(s, c)}$*! := match !*$\nullfun{s}$*! with
  | !*\none*! !*$\hspace{0.9em}$*! !*$\to$*! !*\none*!
  | !*$\some{v}$*! !*$\to$*!
    if !*$c = \nil$*! then !*$\some{v}$*!
    else !*$\fnullable{\plug{v}{c}}$*!
\end{lstlisting}
\end{minipage}

\begin{theorem}
For all LL(1) focused syntax $(s, c)$:
\[
\fnullable{(s, c)} = \nullfun{\unfocus{(s, c)}}
\]
\end{theorem}

\subsection{Zippy Parsing with Derivatives Algorithm}

Using the previous definitions, we can finally present the zippy LL(1) parsing with derivatives algorithm.
Given a focused syntax $(s, c)$ and a token sequence $ts$, the algorithm returns
the value associated with the token sequence, if any.

\begin{minipage}{\linewidth}
\begin{lstlisting}
!*$\fparse{(s, c)}{ts}$*! := match !*$ts$*! with
  | !*$\nil$*! !*$\hspace{1.3em}$*! !*$\to$*! !*$\fnullable{(s, c)}$*!
  | !*$\cons{t}{ts'}$*! !*$\to$*!
    match !*$\fderive{t}{(s, c)}$*! with
    | !*\none*! !*$\hspace{2.9em}$*! !*$\to$*! !*\none*!
    | !*$\some{(s', c')}$*! !*$\to$*! !*$\fparse{(s', c')}{ts'}$*!
\end{lstlisting}
\end{minipage}

\begin{theorem}[Correctness] \label{thm:parse-correct}
For any LL(1) syntax $s$, token sequence $ts$ and value $v$:
\[
\fparse{\focus{s}}{ts} = \some{v} \iff \matches{s}{ts}{v}
\]
\end{theorem}

\subsection{Example Execution}

\Cref{fig:execution}~shows the execution of the algorithm on the syntax for the language $L = \{ \texttt{a}^n \texttt{b}^n\mapsto n\ |\ n \in \mathcal{N} \}$ introduced in~\Cref{sec:formalismexample}, on the sequence of tokens $\sequence{\texttt{a}, \texttt{a}, \texttt{b}, \texttt{b}}$.
The run consists in four $\texttt{derive}$ calls (one per token) and an invocation of $\texttt{result}$.
Each $\texttt{derive}$ call is decomposed into a call to $\texttt{locate}$ and a call to $\texttt{pierce}$.
The focused syntax obtained after each such call is displayed.

\subsection{Runtime Complexity of Parsers}

In this section, we argue that the zippy LL(1) parsing with derivatives algorithm
runs in time linear in the number of tokens (ignoring the cost of user-defined functions appearing in the syntax, which typically apply AST constructors).

A key observation towards this result is that the parsing algorithm never creates new \emph{syntax nodes} apart from trivial $\eps{v}$ nodes.
This can be directly observed by inspecting the code of the parsing algorithm.
This has major implications:
\begin{enumerate}
\item
The computation of the properties of syntaxes by propagator networks can be performed once and for all before processing any input token.
The properties of a syntax can then be directly stored in the syntax node.
The process thus takes constant time with respect to number of input tokens.
\item
Calls to $\texttt{pierce}$ made by the parsing algorithm can only be made on a fixed number of preexisting syntaxes, whose sizes are therefore constant with respect to the input size.
Thus each call runs in constant time with respect to the number of input tokens.
\item
For the same reason, the number of layers added to the context by $\texttt{pierce}$ is constant with respect to the input size.
\end{enumerate}

We now argue that the complexity of the algorithm is proportional to the number of \emph{layer nodes} traversed at parse time.
Indeed, all operations performed by the parsing algorithm (calls to $\texttt{pierce}$, property checks and so on) only take constant time with respect to the number of input tokens and are performed at most once per traversed layer.

We now show that the number of layers traversed at parse time is linear in the input size.
We observe that the $\texttt{plug}$ procedure is the only procedure that directly traverses the stack of layers.
Layer nodes are only ever visited once by $\texttt{plug}$ in their entire lifetime: Indeed, once visited, layers are removed from the context and completely discarded.
Therefore, the number of traversed layers is bounded by the number of layers created at parse time.

We now show that the number of layers created at parse time is linear in the input size.
Only two procedures ever create layers: $\texttt{pierce}$ and $\texttt{plug}$.
\begin{itemize}
\item
As previously argued, $\texttt{pierce}$ is only ever invoked on a finite number of preexisting syntaxes.
For this reason the maximal number of layers nodes that are created by a single call to $\texttt{pierce}$ is constant with respect to the input size.
Since there is at most a single call to $\texttt{pierce}$ per input token, the total number of layers created by $\texttt{pierce}$ is linear in the input size.
\item
Additionally, the $\texttt{plug}$ procedure itself can create layers.
It however does so in a very limited way: $\texttt{plug}$ only creates a $\texttt{prepend}$ layer in case the layer it popped from the context was a $\texttt{follow-by}$ layer.
Since $\texttt{follow-by}$ layers can only be created by $\texttt{pierce}$, the number of layers created by $\texttt{plug}$ is bounded by the number of layers created by $\texttt{pierce}$, and thus is linear in the input size.
\end{itemize}
We therefore have that the total number of layers ever created at parse time is bounded linearly in the input size, which concludes the proof that the zippy LL(1) parsing with derivatives algorithm runs in time linear in the number of input tokens.

\subsection{Connections to Traditional Parsing}

The zippy LL(1) parsing with derivatives algorithm that we have presented in this section shares many features with the traditional LL(1) parsing algorithm.
Immediately, we can observe that both algorithms maintain a stack of rules to be applied on subsequent input.
Interestingly, we arrived at that stack rather naturally by introducing a focus within our syntaxes.
Furthermore, our \texttt{derive} procedure corresponds to the table-based lookup procedure of the traditional algorithm.
Instead of storing the transitions in a table, our transitions are obtained by calling \texttt{pierce} on individual nodes of the syntax.
If we were to pre-compute the layers added by \texttt{pierce} for every kind $k$ in the first set of nodes of syntaxes, we would arrive at an almost identical approach
(with a new and formal proof of correctness).

Additionally, the zippy parsing with derivatives algorithm that we have presented in this section can be adapted to support general context-free expressions, and so by representing the context as a graph instead of a linear stack.
The generalised version of the zippy parsing with derivatives algorithm is reminiscent of the GLL parsing algorithm~\cite{scott2010gll}.

%% file: trees.tex
\tikzset{snode/.style={minimum size=0.25cm, align=center}}
\tikzset{lnode/.style={thin, color=darkgray}}
\tikzset{>=to}
\begin{figure}
\centering
\begin{tikzpicture}[xscale=0.4, yscale=0.45]


\node[lnode] (s1name) at (0, 4) {$s_1$};

\node[snode, label=center:\scriptsize$f \circledcirc$] (s1app1) at (2, 4) {};
\node[snode, label=center:$\cdot$] (s1seq1) at (2, 3) {};
\draw[->] (s1app1) -- (s1seq1);
\node[snode, label=center:$\cdot$] (s1seq2) at (1, 2) {};
\draw[->] (s1seq1) -- (s1seq2);
\node[snode, label=center:\scriptsize$\elem{B}$] (s1elemB1) at (3, 2) {};
\draw[->] (s1seq1) -- (s1elemB1);
\node[snode, label=center:\scriptsize$\eps{\texttt{a}}$] (s1epsa1) at (0, 1) {};
\draw[->] (s1seq2) -- (s1epsa1);
\node[snode, label=center:\scriptsize$\var{x}$] (s1var1) at (2, 1) {};
\draw[->] (s1seq2) -- (s1var1);


\node[lnode] (s2name) at (0 + 6, 4) {$s_2$};

\node[snode, label=center:\scriptsize$f \circledcirc$] (s2app1) at (2 + 6, 4) {};
\node[snode, label=center:$\cdot$] (s2seq1) at (2 + 6, 3) {};
\draw[->] (s2app1) -- (s2seq1);
\node[snode, label=center:$\cdot$] (s2seq2) at (1 + 6, 2) {};
\draw[->] (s2seq1) -- (s2seq2);
\node[snode, label=center:\scriptsize$\elem{B}$] (s2elemB1) at (3 + 6, 2) {};
\draw[->] (s2seq1) -- (s2elemB1);
\node[snode, label=center:\scriptsize$\eps{\texttt{a}}$] (s2epsa1) at (0 + 6, 1) {};
\draw[->] (s2seq2) -- (s2epsa1);

\node[snode, label=center:\scriptsize$f \circledcirc$] (s2app2) at (2 + 6, 1) {};
\draw[->] (s2seq2) -- (s2app2);
\node[snode, label=center:$\cdot$] (s2seq3) at (2 + 6, 0) {};
\draw[->] (s2app2) -- (s2seq3);
\node[snode, label=center:$\cdot$] (s2seq4) at (1 + 6, -1) {};
\draw[->] (s2seq3) -- (s2seq4);
\node[snode, label=center:\scriptsize$\elem{B}$] (s2elemB2) at (3 + 6, -1) {};
\draw[->] (s2seq3) -- (s2elemB2);
\node[snode, label=center:\scriptsize$\eps{\texttt{a}}$] (s2epsa2) at (0 + 6, -2) {};
\draw[->] (s2seq4) -- (s2epsa2);
\node[snode, label=center:\scriptsize$\var{x}$] (s2var2) at (2 + 6, -2) {};
\draw[->] (s2seq4) -- (s2var2);


\node[lnode] (s3name) at (0 + 12, 4) {$s_3$};

\node[snode, label=center:\scriptsize$f \circledcirc$] (s3app1) at (2 + 12, 4) {};
\node[snode, label=center:$\cdot$] (s3seq1) at (2 + 12, 3) {};
\draw[->] (s3app1) -- (s3seq1);
\node[snode, label=center:$\cdot$] (s3seq2) at (1 + 12, 2) {};
\draw[->] (s3seq1) -- (s3seq2);
\node[snode, label=center:\scriptsize$\elem{B}$] (s3elemB1) at (3 + 12, 2) {};
\draw[->] (s3seq1) -- (s3elemB1);
\node[snode, label=center:\scriptsize$\eps{\texttt{a}}$] (s3epsa1) at (0 + 12, 1) {};
\draw[->] (s3seq2) -- (s3epsa1);

\node[snode, label=center:\scriptsize$f \circledcirc$] (s3app2) at (2 + 12, 1) {};
\draw[->] (s3seq2) -- (s3app2);
\node[snode, label=center:$\cdot$] (s3seq3) at (2 + 12, 0) {};
\draw[->] (s3app2) -- (s3seq3);
\node[snode, label=center:$\cdot$] (s3seq4) at (1 + 12, -1) {};
\draw[->] (s3seq3) -- (s3seq4);
\node[snode, label=center:\scriptsize$\elem{B}$] (s3elemB2) at (3 + 12, -1) {};
\draw[->] (s3seq3) -- (s3elemB2);
\node[snode, label=center:\scriptsize$\eps{\texttt{a}}$] (s3epsa2) at (0 + 12, -2) {};
\draw[->] (s3seq4) -- (s3epsa2);

\node[snode, label=center:\scriptsize$f \circledcirc$] (s3app3) at (2 + 12, -2) {};
\draw[->] (s3seq4) -- (s3app3);
\node[snode, label=center:$\cdot$] (s3seq5) at (2 + 12, -3) {};
\draw[->] (s3app3) -- (s3seq5);
\node[snode, label=center:$\cdot$] (s3seq6) at (1 + 12, -4) {};
\draw[->] (s3seq5) -- (s3seq6);
\node[snode, label=center:\scriptsize$\elem{B}$] (s3elemB3) at (3 + 12, -4) {};
\draw[->] (s3seq5) -- (s3elemB3);
\node[snode, label=center:\scriptsize$\eps{\texttt{a}}$] (s3epsa3) at (0 + 12, -5) {};
\draw[->] (s3seq6) -- (s3epsa3);
\node[snode, label=center:\scriptsize$\var{x}$] (s3var3) at (2 + 12, -5) {};
\draw[->] (s3seq6) -- (s3var3);

\end{tikzpicture}
\caption{Representation of the syntaxes $s_1$, $s_2$ and $s_3$.}
\label{fig:syntaxtrees}
\end{figure}

%% file: focusedtree.tex

\tikzset{layernode/.style={minimum size=0.25cm, align=center}}
\tikzset{arr/.style={thin, ->, dashed, shorten >=0.2cm}}
\centering
\begin{tikzpicture}[xscale=0.4, yscale=0.45]
\node[layernode, label=center:\scriptsize$f \circledcirc$] (app1) at (2, 4) {};
\node[layernode, label=center:$\cdot$] (seq1) at (2, 3) {};
\draw[<-, thick] (app1) -- (seq1);
\node[layernode, label=center:$\cdot$] (seq2) at (1, 2) {};
\draw[<-, thick] (seq1) -- (seq2);
\node[snode, label=center:\scriptsize$\elem{B}$] (elemB1) at (3, 2) {};
\draw[->] (seq1) -- (elemB1);
\node[snode, label=center:\scriptsize$\eps{\texttt{a}}$] (epsa1) at (0, 1) {};
\draw[->] (seq2) -- (epsa1);

\node[layernode, label=center:\scriptsize$f \circledcirc$] (app2) at (2, 1) {};
\draw[<-, thick] (seq2) -- (app2);
\node[layernode, label=center:$\cdot$] (seq3) at (2, 0) {};
\draw[<-, thick] (app2) -- (seq3);
\node[layernode, label=center:$\cdot$] (seq4) at (1, -1) {};
\draw[<-, thick] (seq3) -- (seq4);
\node[snode, label=center:\scriptsize$\elem{B}$] (elemB2) at (3, -1) {};
\draw[->] (seq3) -- (elemB2);
\node[snode, label=center:\scriptsize$\eps{\texttt{a}}$] (epsa2) at (0, -2) {};
\draw[->] (seq4) -- (epsa2);

\node[layernode, label=center:\scriptsize$f \circledcirc$] (app3) at (2, -2) {};
\draw[<-, thick] (seq4) -- (app3);
\node[layernode, label=center:$\cdot$] (seq5) at (2, -3) {};
\draw[<-, thick] (app3) -- (seq5);
\node[layernode, label=center:$\cdot$] (seq6) at (1, -4) {};
\draw[<-, thick] (seq5) -- (seq6);
\node[snode, label=center:\scriptsize$\elem{B}$] (elemB3) at (3, -4) {};
\draw[->] (seq5) -- (elemB3);
\node[snode, label=center:\scriptsize$\eps{\texttt{a}}$] (epsa3) at (0, -5) {};
\draw[<-, thick] (seq6) -- (epsa3);
\node[snode, label=center:\scriptsize$\var{x}$] (var3) at (2, -5) {};
\draw[->] (seq6) -- (var3);

\node[black] (layerslabel) at (-5, -1) {\small{Layers}};

\node[black] (focuslabel) at (-5, -5) {\small{Focal point}};

\path
    (layerslabel) edge[arr, bend left] (app1)
    (layerslabel) edge[arr, bend left] (seq1)
    (layerslabel) edge[arr, bend left] (seq2)
     (layerslabel) edge[arr] (app2)
    (layerslabel) edge[arr] (seq3)
    (layerslabel) edge[arr] (seq4)
     (layerslabel) edge[arr, bend right] (app3)
    (layerslabel) edge[arr, bend right] (seq5)
    (layerslabel) edge[arr, bend right] (seq6)
    (focuslabel) edge[arr] (epsa3);

\end{tikzpicture}

%% file: execution.tex

\tikzset{snode/.style={minimum size=0.28cm, align=center}}
\tikzset{lnode/.style={thin, color=darkgray}}
\tikzset{back/.style={<-, thick}}
\tikzset{prog/.style={->, shorten >=0.1cm, shorten <=0.1cm}}
\tikzset{prog2/.style={->, shorten >=0.1cm, shorten <=0.1cm}}
\tikzset{>=to}

\begin{tikzpicture}[xscale=0.35, yscale=0.45]


\draw[prog] (-5 + 1 * 5, 7) to node[midway, below] {\scriptsize\texttt{locate}} (0 + 1 * 5, 7);
\draw[prog] (-5 + 2 * 5, 7) to node[midway, below] {\scriptsize\texttt{pierce}} (0 + 2 * 5, 7);
\draw[prog] (-5 + 3 * 5, 7) to node[midway, below] {\scriptsize\texttt{locate}} (0 + 3 * 5, 7);
\draw[prog] (-5 + 4 * 5, 7) to node[midway, below] {\scriptsize\texttt{pierce}} (0 + 4 * 5, 7);
\draw[prog] (-5 + 5 * 5, 7) to node[midway, below] {\scriptsize\texttt{locate}} (0 + 5 * 5, 7);
\draw[prog] (-5 + 6 * 5, 7) to node[midway, below] {\scriptsize\texttt{pierce}} (0 + 6 * 5, 7);
\draw[prog] (-5 + 7 * 5, 7) to node[midway, below] {\scriptsize\texttt{locate}} (0 + 7 * 5, 7);
\draw[prog] (-5 + 8 * 5, 7) to node[midway, below] {\scriptsize\texttt{pierce}} (0 + 8 * 5, 7);

\draw[prog2] (-10 + 1 * 10, 7.3) to node[midway, above] {\scriptsize{derivation by \texttt{a}}} (0 + 1 * 10, 7.3);
\draw[prog2] (-10 + 2 * 10, 7.3) to node[midway, above] {\scriptsize{derivation by \texttt{a}}} (0 + 2 * 10, 7.3);
\draw[prog2] (-10 + 3 * 10, 7.3) to node[midway, above] {\scriptsize{derivation by \texttt{b}}} (0 + 3 * 10, 7.3);
\draw[prog2] (-10 + 4 * 10, 7.3) to node[midway, above] {\scriptsize{derivation by \texttt{b}}} (0 + 4 * 10, 7.3);

\draw[prog2] (-10 + 5 * 10, 7.3) to node[midway, above] {\scriptsize\texttt{result}} (- 5 + 5 * 10, 7.3);


\node[snode, label=center:\scriptsize$\var{x}$] (t0var1) at (0, 6) {};


\node[snode, label=center:\scriptsize$\var{x}$] (t1var1) at (0 + 1 * 5, 6) {};


\node[snode, label=center:\scriptsize$f \circledcirc$] (t2app1) at (0 + 2 * 5, 6) {};
\node[snode, label=center:$\cdot$] (t2seq1) at (0 + 2 * 5, 5) {};
\draw[back] (t2app1) -- (t2seq1);
\node[snode, label=center:$\cdot$] (t2seq2) at (-1 + 2 * 5, 4) {};
\draw[back] (t2seq1) -- (t2seq2);
\node[snode, label=center:\scriptsize$\elem{B}$] (t2elemB1) at (1  + 2 * 5, 4) {};
\draw[->] (t2seq1) -- (t2elemB1);
\node[snode, label=center:\scriptsize$\eps{\texttt{a}}$] (t2epsa1) at (-2 + 2 * 5, 3) {};
\draw[back] (t2seq2) -- (t2epsa1);
\node[snode, label=center:\scriptsize$\var{x}$] (t2var1) at (0 + 2 * 5, 3) {};
\draw[->] (t2seq2) -- (t2var1);


\node[snode, label=center:\scriptsize$f \circledcirc$] (t3app1) at (0 + 3 * 5, 6) {};
\node[snode, label=center:$\cdot$] (t3seq1) at (0 + 3 * 5, 5) {};
\draw[back] (t3app1) -- (t3seq1);
\node[snode, label=center:$\cdot$] (t3seq2) at (-1 + 3 * 5, 4) {};
\draw[back] (t3seq1) -- (t3seq2);
\node[snode, label=center:\scriptsize$\elem{B}$] (t3elemB1) at (1  + 3 * 5, 4) {};
\draw[->] (t3seq1) -- (t3elemB1);
\node[snode, label=center:\scriptsize$\eps{\texttt{a}}$] (t3epsa1) at (-2 + 3 * 5, 3) {};
\draw[->] (t3seq2) -- (t3epsa1);
\node[snode, label=center:\scriptsize$\var{x}$] (t3var1) at (0 + 3 * 5, 3) {};
\draw[back] (t3seq2) -- (t3var1);


\node[snode, label=center:\scriptsize$f \circledcirc$] (t4app1) at (0 + 4 * 5, 6) {};
\node[snode, label=center:$\cdot$] (t4seq1) at (0 + 4 * 5, 5) {};
\draw[back] (t4app1) -- (t4seq1);
\node[snode, label=center:$\cdot$] (t4seq2) at (-1 + 4 * 5, 4) {};
\draw[back] (t4seq1) -- (t4seq2);
\node[snode, label=center:\scriptsize$\elem{B}$] (t4elemB1) at (1  + 4 * 5, 4) {};
\draw[->] (t4seq1) -- (t4elemB1);
\node[snode, label=center:\scriptsize$\eps{\texttt{a}}$] (t4epsa1) at (-2 + 4 * 5, 3) {};
\draw[->] (t4seq2) -- (t4epsa1);

\node[snode, label=center:\scriptsize$f \circledcirc$] (t4app2) at (0 + 4 * 5, 6 - 3) {};
\draw[back] (t4seq2) -- (t4app2);
\node[snode, label=center:$\cdot$] (t4seq3) at (0 + 4 * 5, 5 - 3) {};
\draw[back] (t4app2) -- (t4seq3);
\node[snode, label=center:$\cdot$] (t4seq4) at (-1 + 4 * 5, 4 - 3) {};
\draw[back] (t4seq3) -- (t4seq4);
\node[snode, label=center:\scriptsize$\elem{B}$] (t4elemB2) at (1  + 4 * 5, 4 - 3) {};
\draw[->] (t4seq3) -- (t4elemB2);
\node[snode, label=center:\scriptsize$\eps{\texttt{a}}$] (t4epsa2) at (-2 + 4 * 5, 3 - 3) {};
\draw[back] (t4seq4) -- (t4epsa2);
\node[snode, label=center:\scriptsize$\var{x}$] (t4var1) at (0 + 4 * 5, 3 - 3) {};
\draw[->] (t4seq4) -- (t4var1);


\node[snode, label=center:\scriptsize$f \circledcirc$] (t5app1) at (0 + 5 * 5, 6) {};
\node[snode, label=center:$\cdot$] (t5seq1) at (0 + 5 * 5, 5) {};
\draw[back] (t5app1) -- (t5seq1);
\node[snode, label=center:$\cdot$] (t5seq2) at (-1 + 5 * 5, 4) {};
\draw[back] (t5seq1) -- (t5seq2);
\node[snode, label=center:\scriptsize$\elem{B}$] (t5elemB1) at (1  + 5 * 5, 4) {};
\draw[->] (t5seq1) -- (t5elemB1);
\node[snode, label=center:\scriptsize$\eps{\texttt{a}}$] (t5epsa1) at (-2 + 5 * 5, 3) {};
\draw[->] (t5seq2) -- (t5epsa1);

\node[snode, label=center:\scriptsize$f \circledcirc$] (t5app2) at (0 + 5 * 5, 6 - 3) {};
\draw[back] (t5seq2) -- (t5app2);
\node[snode, label=center:$\cdot$] (t5seq3) at (0 + 5 * 5, 5 - 3) {};
\draw[back] (t5app2) -- (t5seq3);
\node[snode, label=center:\scriptsize$\eps{(\texttt{a}, 0)}$] (t5epsacc) at (-1 + 5 * 5, 4 - 3) {};
\draw[->] (t5seq3) -- (t5epsacc);
\node[snode, label=center:\scriptsize$\elem{B}$] (t5elemB2) at (1  + 5 * 5, 4 - 3) {};
\draw[back] (t5seq3) -- (t5elemB2);


\node[snode, label=center:\scriptsize$f \circledcirc$] (t6app1) at (0 + 6 * 5, 6) {};
\node[snode, label=center:$\cdot$] (t6seq1) at (0 + 6 * 5, 5) {};
\draw[back] (t6app1) -- (t6seq1);
\node[snode, label=center:$\cdot$] (t6seq2) at (-1 + 6 * 5, 4) {};
\draw[back] (t6seq1) -- (t6seq2);
\node[snode, label=center:\scriptsize$\elem{B}$] (t6elemB1) at (1  + 6 * 5, 4) {};
\draw[->] (t6seq1) -- (t6elemB1);
\node[snode, label=center:\scriptsize$\eps{\texttt{a}}$] (t6epsa1) at (-2 + 6 * 5, 3) {};
\draw[->] (t6seq2) -- (t6epsa1);

\node[snode, label=center:\scriptsize$f \circledcirc$] (t6app2) at (0 + 6 * 5, 6 - 3) {};
\draw[back] (t6seq2) -- (t6app2);
\node[snode, label=center:$\cdot$] (t6seq3) at (0 + 6 * 5, 5 - 3) {};
\draw[back] (t6app2) -- (t6seq3);
\node[snode, label=center:\scriptsize$\eps{(\texttt{a}, 0)}$] (t6epsacc) at (-1 + 6 * 5, 4 - 3) {};
\draw[->] (t6seq3) -- (t6epsacc);
\node[snode, label=center:\scriptsize$\eps{\texttt{b}}$] (t6epsb) at (1  + 6 * 5, 4 - 3) {};
\draw[back] (t6seq3) -- (t6epsb);


\node[snode, label=center:\scriptsize$f \circledcirc$] (t7app1) at (0 + 7 * 5, 6) {};
\node[snode, label=center:$\cdot$] (t7seq1) at (0 + 7 * 5, 5) {};
\draw[back] (t7app1) -- (t7seq1);
\node[snode, label=center:\scriptsize$\eps{(\texttt{a}, 1)}$] (t7epsacc) at (-1 + 7 * 5, 4) {};
\draw[->] (t7seq1) -- (t7epsacc);
\node[snode, label=center:\scriptsize$\elem{B}$] (t7elemB1) at (1  + 7 * 5, 4) {};
\draw[back] (t7seq1) -- (t7elemB1);


\node[snode, label=center:\scriptsize$f \circledcirc$] (t8app1) at (0 + 8 * 5, 6) {};
\node[snode, label=center:$\cdot$] (t8seq1) at (0 + 8 * 5, 5) {};
\draw[back] (t8app1) -- (t8seq1);
\node[snode, label=center:\scriptsize$\eps{(\texttt{a}, 1)}$] (t8epsacc) at (-1 + 8 * 5, 4) {};
\draw[->] (t8seq1) -- (t8epsacc);
\node[snode, label=center:\scriptsize$\eps{\texttt{b}}$] (t8epsb) at (1  + 8 * 5, 4) {};
\draw[back] (t8seq1) -- (t8epsb);


\node[snode, label=center:$\some{2}$] (t9res) at (0 + 9 * 5, 6) {};

\end{tikzpicture}

%% file: coq-proofs.tex

\section{Coq Proofs}\label{sec:coq}

We formalised the parsing with derivatives algorithm with zippy syntaxes in Coq
(around 9000 lines).
The Coq proofs are freely available at \url{https://github.com/epfl-lara/scallion-proofs}.
We defined the recursive functions that require non-trivial
measures using the Equations library~\cite{sozeau2019equations}. There are two main parts in the
formalism: one (around 7000 lines, including around 3000 lines about propagator
networks constructions and proofs) to define the functions corresponding to the basic properties of syntaxes and many properties about them,
and one (around 2000 lines) for the parsing algorithm based on zippy syntaxes (and its
correctness).

In the first part, we defined for each function the inductive rules as described
in \Cref{fig:defs} and a corresponding propagator network that gives a way to
compute the function. We defined a uniform way to specify these rules on
syntaxes using the notion of a \emph{description}.
We then made a generic construction that takes a syntax and a description, and
builds a propagator network that computes the function corresponding to the
description on the syntax. This propagator network has one cell per node in the
syntax, and each cell is updated using the inductive rules based on the cells
corresponding to the children of the syntax. We proved soundness and
completeness of this construction.
Our Coq definitions of propagator networks (and their termination guarantees)
are general and can be reused independently of this paper and independently of
syntaxes.

In the second part, we defined zippy syntaxes, the functions \texttt{plug},
\texttt{locate}, \texttt{pierce}, \texttt{derive} and proved all the necessary
properties to show the correctness of parsing as stated in
\Cref{thm:parse-correct}. In particular, we proved that these functions
terminate, that they do not introduce conflicts, and that they produce syntaxes
that recognise the expected languages (\Cref{thm:derive-lang}).

%% file: table-performance.tex

\begin{table*}[htbp]
\caption{Performance comparison between simple LL(1) parsing with derivatives (Simple), zippy LL(1) parsing with derivatives (Zippy), and Scala Parser Combinators (SPC) for parsing JSON. Entries marked with \dag\ encountered a stack overflow. Entries correspond to the mean of 36 measurements on a hot JVM. }
\begin{center}
\begin{tabular}{|C{2cm}|C{2cm}||C{1.5cm}|C{1.5cm}|C{1.5cm}||C{1.5cm}|C{1.5cm}|C{1.5cm}|}
\hline
\multirow{2}{*}{File size (KB)} & \multirow{2}{*}{Tokens} & \multicolumn{3}{c||}{Parse time (ms)} & \multicolumn{3}{c|}{Speed (token/ms)} \\
\cline{3-8}
 &  & Simple & Zippy & SPC & Simple & Zippy & SPC \\
\hline
100 & 9649 & 99.9 & 2.8 & 2.3 & 96.6 & 3446.0 & 4195.2\\
1000 & 97821 & 7069.2 & 14.3 & 19.0 & 13.8 & 6840.6 & 5159.3\\
10000 & 971501 & \dag & 150.2 & 166.0 & \dag & 6468.0 & 5852.4\\
\hline
\end{tabular}
\end{center}
\label{tab:perf}
\end{table*}

%% file: implementation.tex

\section{Parsing and Printing Combinators}\label{sec:impl}

In this section, we discuss the implementation of syntaxes as a parsing and printing combinators framework in Scala.
The framework is freely available under an open source license\footnote{The framework is available at \url{https://github.com/epfl-lara/scallion}}.
The Scala implementation closely follows the Coq formalism.
For performance reasons, we did not mechanically extract an implementation from the formalisation.

\subsection{Syntax Definition}
\label{sec:adt}

Syntaxes are defined as a generalised algebraic datatype named \texttt{Syntax[A]}.
Each construct of the formalism straightforwardly corresponds to one constructor of the datatype.
The environment syntaxes are directly stored in the instance of the corresponding variable syntaxes.
To enable (mutually) recursive syntaxes, the syntax $\getdef{x}$ is stored in a \emph{lazy} field of the instance of $\var{x}$.

\subsection{Computing Properties of Syntaxes}

Properties of syntaxes (productivity, nullability, first sets etc.) are stored as public fields of \texttt{Syntax} instances.
Fields are used by the LL(1) checking procedure and by the parsing algorithm.
In addition, the first set of a syntax can be used to suggest fixes in case of parse errors.
Propagator networks~\cite{steele1980definition, radul2009propagation} are used to initialise the fields ahead of parsing, as explained in~\Cref{sec:networks}.

The LL(1) property of syntaxes can be checked via a simple method call.
In case a syntax is not LL(1), the list of conflicts can be obtained and their root causes identified.
Coupled with the enumeration capabilities of the framework, users of the framework can easily get examples of token sequences which lead to  conflicts.

\subsection{Parsing}

Parsing is performed via the \texttt{apply} method of \texttt{Syntax[A]}.
The method takes as input an \texttt{Iterator} of tokens and returns a value of type \texttt{ParseResult[A]}, which can be:
\begin{enumerate}
\item
\texttt{Parsed(value, descr)}, which indicates that the given \texttt{value} (of type \texttt{A}) was successfully parsed.
\item
\texttt{UnexpectedToken(token, descr)}, indicating that \texttt{token} was not expected. Values from the input iterator are not consumed beyond that token.
\item
\texttt{UnexpectedEnd(descr)}, which indicates that the end of input was not expected.
\end{enumerate}
In each case, a residual focused syntax \texttt{descr} is also returned. This syntax represents the state at the end of parsing or at an error point.
This syntax can be queried and used as any other syntax.
In particular, it can be used for error reporting and recovery, and to resume parsing.
Thanks to the derivatives-based algorithm, this syntax is available "for free".

The framework faithfully implements the zippy LL(1) parsing with derivatives presented in~\Cref{sec:zippy}.
The methods \texttt{plug}, \texttt{locate} and \texttt{pierce} are tail-recursive, which ensures that the call stack of underlying virtual machine does not overflow during parsing.
The framework also implements memoisation of calls to \texttt{pierce}.
The additional layers of context returned by \texttt{pierce} are stored in reverse order for fast concatenation.

\subsection{Enumeration and Pretty Printing}

Our framework also supports 1) enumeration of recognised sequences of token kinds and 2) pretty printing, that is, the enumeration of token sequences that would be parsed into given values.
To support this second feature, the constructor for $\map{f}{s}$ accepts an extra argument for the inverse of the function to be applied on produced values.
Whenever local inverses are correct, all generated pretty printed sequences are guaranteed to parse and generate a given value.
For both enumeration and pretty printing, sequences are produced in order of increasing length, typically resulting in the first having, e.g., the fewest number of parentheses. 

\subsection{Library of Combinators}

A library of useful combinators is offered to programmers, such as repetition combinators (\texttt{many}, \texttt{many1}), repetition with separators combinators (\texttt{repsep}, \texttt{rep1sep}), optional combinator (\texttt{opt}), tagged disjunctions (infix method \texttt{||}) and many others. Higher level combinators, such as combinators for infix operators with multiple priority levels and associativities are also available in the library.
All combinators are expressed in terms of the primitive syntaxes and combinators shown in~\Cref{sec:adt}, and have support for pretty printing out of the box.

%% file: evaluation.tex

\section{Experimental Evaluation}

We compare the performance of the presented zippy LL(1) parsing with derivatives algorithm with the simple (non-zippy) LL(1) parsing with derivatives and with Scala Parser Combinators~\cite{scalaparsercombinators}.
The latter is a widely adopted parser combinators library in Scala, which uses recursive descent parsing by default, but also supports packrat parsing.

\Cref{tab:perf}~shows the performance of the three approaches for parsing JSON files of size ranging from 100KB to 10MB.
Each JSON file contains a single large array of objects, each containing several string and array fields.
The JSON files were randomly generated using an online JSON generator~\cite{JSONGenerator}.
The benchmarks were run on a MacBook Pro with Core i7 CPU@2.2GHz and 16 GB RAM,
running Scala 2.12.8 and Java 1.8 on the HotSpot\texttrademark\ JVM.
We used ScalaMeter \cite{Scalameter} as the benchmarking tool.
All three approaches were given tokens from the same lexer. Lexing time is not reported.
The table reports the mean values of 36 measurements.

The zippy LL(1) parsing with derivatives outperforms the simple variant by orders of magnitude.
The speed of the simple LL(1) parsing with derivatives algorithm degrades with the number of tokens,
unlike the speed of the zippy variant.
Moreover, the simple parsing algorithm encounters a stack overflow on large files.

The performance of the zippy LL(1) parsing with derivatives is comparable to the performance of the recursive descent algorithm implemented by the Scala Parser Combinators library.
Worth noting, the zippy LL(1) parsing with derivatives algorithm doesn't suffer from stack overflows, which can occur with recursive descent when parsing deeply nested structures.
Since parsers are often exposed to user inputs, an attacker could exploit this vulnerability in approaches based on recursive descent to cause crashes, and so with a relatively small input JSON file (as small as 2616 bytes in our tests).
Our implementation also offers more comprehensive error
reporting and recovery, in part thanks to encoding of parser states as explicit analysable expressions. 

We also benchmarked the performance of Parseback~\cite{parseback}, a recent Scala implementation of the parsing with derivatives algorithm~\cite{might2011parsing} by one of the original authors, with performance optimisations from~\cite{adams2016complexity}. The results are not reported in~\Cref{tab:perf} as the parser encounters a stack overflow in each of the benchmarks.
The largest file we managed to parse with that library was 1387 bytes long, and it took 1388ms.

\begin{table}[H]
\caption{Performance of the zippy parsing with derivatives algorithm combined with a handwritten lexer (Ours) compared to an ANTLR-generated lexer and parser (ANTLR).}
\begin{center}
\begin{tabular}{|C{2cm}|C{2cm}||C{1.45cm}|C{1.45cm}|}
\hline
\multirow{2}{*}{File size (KB)} & \multirow{2}{*}{Tokens}  & \multicolumn{2}{c|}{Lex \& parse time (ms)} \\
\cline{3-4}
 & &  Ours & ANTLR \\
\hline
100 &  9649 & 6.4 & 1.9\\
1000 & 97821 & 40.9 & 15.8\\
10000 & 971501 & 449.9 & 145.9\\
\hline
\end{tabular}
\end{center}
\label{tab:perfantlr}
\end{table}

Finally, we benchmarked the performance of our approach compared to an ANTLR-generated JSON parser~\cite{parr2013definitive,DBLP:conf/pldi/ParrF11,ANTLRexamples}.
Results are presented in~\Cref{tab:perfantlr}.
Our approach, which doesn't resort to code generation, is a constant factor (\textasciitilde 3) slower than the ANTLR-generated solution.
However, our approach offers a highly flexible and extensible parser combinators interface embedded in a rich programming language, while ANTLR grammars are described in a domain-specific language of limited expressivity.
In addition, while our approach directly builds values of the appropriate type, extra work must be done to convert parse trees produced by the ANTLR-generated parser into proper user-defined JSON values. This extra work is not reported in the results.

In addition to the JSON parser, we have developed parsers for several other non-trivial languages.
We used the presented framework to build a parser and pretty printer for a first-order logic formulas quasiquoter, a parser and pretty printer for lambda-calculus, a parser for an expression language with infix, prefix and postfix operators, as well as several other examples.
We have also used our framework in EPFL CS-320, a third-year BSc compiler construction course with over 40 students.
As a semester-long project, students build a compiler for a subset of Scala.
Students successfully used the presented framework to build their parsers, appreciating the debugging capabilities offered by the framework.

%% file: related.tex

\section{Related Work}

\citet{ford2002packrat} presents \emph{packrat} parsing, a parsing technique for \emph{parsing expression grammars} (PEGs).
Packrat parsers are non-ambiguous and guaranteed to run in linear time through heavy use of memoisation but
tend to be slower than many other linear-time parsing techniques~\cite{becket2008dcgs+, grimm2004practical}.
Whereas PEGs disallow ambiguities through biased choices, LL(1) approaches such as ours support
detecting ambiguities before parsing starts.
We believe that it is better to detect and report ambiguities rather than to hide them.
Our combinators also enjoy more natural algebraic properties, with
our disjunctions being commutative and associative, which is not the case in PEGs, making
the composition of PEGs trickier.

\citet{DBLP:conf/uss/RamananandroDFS19} demonstrate the importance of parsers in security and present combinators for building verified high-performance parser for 
\emph{lower-level} encodings of data formats. In contrast, we
focus on parsing generalisations of context-free grammars.
Formally verified parsers are of special interest to verified compilers such as CompCert~\cite{leroy2009formal} and CakeML~\cite{kumar2014cakeml}.
\citet{koprowski2010trx}~present a formally verified Coq parser interpreter for PEGs.
In recent work authors~\citet{lasser2019verified} present a Coq-verified LL(1) parser generator.
The generated parser uses the traditional table-based LL(1) algorithm, and relies on fix-point computations for properties such as nullability, first sets and others.
While these works operate at the level PEGs or context-free grammars, our work
works on \emph{value-aware} context-free expressions.
As an alternative approach, \citet{jourdan2012validating} developed
a validator (implemented and verified in Coq) for LR(1) parsers.
Their approach works by verifying a posteriori that an automaton-based parser faithfully implements a context-free grammar, while we present a general correctness proof of a parser
operating directly on context-free expressions.
\citet{swierstra1996deterministic} propose parser combinators for LL(1) languages.
Due to their approach based on a shallow embedding of combinators, they are unable to check for LL(1) conflicts a priori.
The parsing procedure they use is based on lookup tables, as opposed to our parsing approach based on derivatives.

Our implementation supports mutually inverse parsing and pretty printing, which is also present in \citet{rendel2010invertible} based on \emph{syntactic descriptions} and using recursive descent parsing (instead of using derivatives).

\citet{Krishnaswami:2019:TAA:3314221.3314625} propose a type-system for LL(1) context-free expressions.
They use the usual conversion to push-down automata for parsing, and rely on code-generation for good performance.
In their approach, the various properties of context-free expressions (nullability, first sets, etc.) are obtained via fixpoint computations, as opposed to our approach based on propagator networks.
They use a weaker definition of \emph{should-not-follow} set (which they call \emph{follow-last} set, abbreviated as \textsc{FLast}).
Their type system is more restrictive than ours as it does not allow nullable expressions to appear on the left of sequences.

\citet{might2011parsing} present a parsing algorithm for context-free expressions based on derivatives.
Compared to our paper, their approach is not restricted to only LL(1) expressions, but is applicable to a wider family of context-free expressions.
The worse-case complexity of their approach is cubic in general~\cite{adams2016complexity}, and can be shown to be (at least) quadratic for LL(1) expressions by following an argument similar to \Cref{sec:inefficient}. Our approach is limited to LL(1) languages but has guaranteed linear time complexity thanks to the use of a zipper-like data structure.

\citet{Brachthauser:2016:PFD:2983990.2984026} showcase how derivatives can be used to augment the language of parser combinators and gain fine-grained control over the input stream.
The \texttt{feed} and \texttt{done} combinators they introduce can be straightforwardly implemented as derived combinators in our setting.
However, most of the examples and patterns demonstrating the power of their approach require a monadic \texttt{flatMap} combinator, which we do not support.

\citet{Henriksen:2019:DGS:3366395.3360553}~show a parsing technique based on derivatives for context-free grammars.
They show that their approach is equivalent to Earley's algorithm~\cite{earley1970efficient} and
argue that parsing with derivatives has deep connections with traditional parsing techniques.
In this paper, we reinforce such connection, linking traditional LL(1) parsing to efficient
parsing with derivatives.

%% file: acknowledgements.tex

\begin{acks}
  This work is supported by the Swiss National Science Foundation Project number 200021\_175676 as well as EPFL.
We thank the anonymous PLDI'20 and PLDI'20 Artifact Evaluation reviewers for their thorough and insightful reviews.
The authors would also like to thank Joachim Hugonot, Maxime Kjaer, 
Dragana Milovan\v{c}evi\'{c}, Romain Ruetschi, Georg Schmid, and Nataliia Stulova for their feedback and help on this work.
\end{acks}

%% file: paper.bbl

\begin{thebibliography}{59}


\ifx \showCODEN    \undefined \def \showCODEN     #1{\unskip}     \fi
\ifx \showDOI      \undefined \def \showDOI       #1{#1}\fi
\ifx \showISBNx    \undefined \def \showISBNx     #1{\unskip}     \fi
\ifx \showISBNxiii \undefined \def \showISBNxiii  #1{\unskip}     \fi
\ifx \showISSN     \undefined \def \showISSN      #1{\unskip}     \fi
\ifx \showLCCN     \undefined \def \showLCCN      #1{\unskip}     \fi
\ifx \shownote     \undefined \def \shownote      #1{#1}          \fi
\ifx \showarticletitle \undefined \def \showarticletitle #1{#1}   \fi
\ifx \showURL      \undefined \def \showURL       {\relax}        \fi
\providecommand\bibfield[2]{#2}
\providecommand\bibinfo[2]{#2}
\providecommand\natexlab[1]{#1}
\providecommand\showeprint[2][]{arXiv:#2}

\bibitem[\protect\citeauthoryear{Adams, Hollenbeck, and Might}{Adams
  et~al\mbox{.}}{2016}]%
        {adams2016complexity}
\bibfield{author}{\bibinfo{person}{Michael~D. Adams}, \bibinfo{person}{Celeste
  Hollenbeck}, {and} \bibinfo{person}{Matthew Might}.}
  \bibinfo{year}{2016}\natexlab{}.
\newblock \showarticletitle{On the Complexity and Performance of Parsing with
  Derivatives}. In \bibinfo{booktitle}{\emph{Proceedings of the 37th ACM
  SIGPLAN Conference on Programming Language Design and Implementation}} (Santa
  Barbara, CA, USA) \emph{(\bibinfo{series}{PLDI '16})}.
  \bibinfo{publisher}{ACM}, \bibinfo{address}{New York, NY, USA},
  \bibinfo{pages}{224--236}.
\newblock
\showISBNx{978-1-4503-4261-2}
\urldef\tempurl%
\url{https://doi.org/10.1145/2908080.2908128}
\showDOI{\tempurl}


\bibitem[\protect\citeauthoryear{Aho, Lam, Sethi, and Ullman}{Aho
  et~al\mbox{.}}{2006}]%
        {Aho:2006:CPT:1177220}
\bibfield{author}{\bibinfo{person}{Alfred~V. Aho}, \bibinfo{person}{Monica~S.
  Lam}, \bibinfo{person}{Ravi Sethi}, {and} \bibinfo{person}{Jeffrey~D.
  Ullman}.} \bibinfo{year}{2006}\natexlab{}.
\newblock \bibinfo{booktitle}{\emph{Compilers: Principles, Techniques, and
  Tools (2nd Edition)}}.
\newblock \bibinfo{publisher}{Addison-Wesley Longman Publishing Co., Inc.},
  \bibinfo{address}{Boston, MA, USA}.
\newblock
\showISBNx{0321486811}


\bibitem[\protect\citeauthoryear{Aho and Ullman}{Aho and Ullman}{1972}]%
        {DBLP:books/lib/AhoU72}
\bibfield{author}{\bibinfo{person}{Alfred~V. Aho} {and}
  \bibinfo{person}{Jeffrey~D. Ullman}.} \bibinfo{year}{1972}\natexlab{}.
\newblock \bibinfo{booktitle}{\emph{The theory of parsing, translation, and
  compiling. 1: Parsing}}.
\newblock \bibinfo{publisher}{Prentice-Hall}.
\newblock
\showISBNx{0-13-914556-7}


\bibitem[\protect\citeauthoryear{Ausaf, Dyckhoff, and Urban}{Ausaf
  et~al\mbox{.}}{2016}]%
        {Posix-Lexing-AFP}
\bibfield{author}{\bibinfo{person}{Fahad Ausaf}, \bibinfo{person}{Roy
  Dyckhoff}, {and} \bibinfo{person}{Christian Urban}.}
  \bibinfo{year}{2016}\natexlab{}.
\newblock \showarticletitle{POSIX Lexing with Derivatives of Regular
  Expressions}.
\newblock \bibinfo{journal}{\emph{Archive of Formal Proofs}}
  (\bibinfo{date}{May} \bibinfo{year}{2016}).
\newblock
\showISSN{2150-914x}
\newblock
\shownote{\url{http://isa-afp.org/entries/Posix-Lexing.html}, Formal proof
  development.}


\bibitem[\protect\citeauthoryear{Becket and Somogyi}{Becket and
  Somogyi}{2008}]%
        {becket2008dcgs+}
\bibfield{author}{\bibinfo{person}{Ralph Becket} {and} \bibinfo{person}{Zoltan
  Somogyi}.} \bibinfo{year}{2008}\natexlab{}.
\newblock \showarticletitle{DCGs+ memoing= packrat parsing but is it worth
  it?}. In \bibinfo{booktitle}{\emph{International Symposium on Practical
  Aspects of Declarative Languages}}. Springer, \bibinfo{pages}{182--196}.
\newblock


\bibitem[\protect\citeauthoryear{Blog}{Blog}{2019}]%
        {CloudflareBug}
\bibfield{author}{\bibinfo{person}{Cloudflare Blog}.}
  \bibinfo{year}{2019}\natexlab{}.
\newblock \bibinfo{title}{Incident report on memory leak caused by {Cloudflare}
  parser bug}.
\newblock
  \bibinfo{howpublished}{\url{https://blog.cloudflare.com/incident-report-on-memory-leak-caused-by-cloudflare-parser-bug/}}.
\newblock


\bibitem[\protect\citeauthoryear{Brachth\"{a}user, Rendel, and
  Ostermann}{Brachth\"{a}user et~al\mbox{.}}{2016}]%
        {Brachthauser:2016:PFD:2983990.2984026}
\bibfield{author}{\bibinfo{person}{Jonathan~Immanuel Brachth\"{a}user},
  \bibinfo{person}{Tillmann Rendel}, {and} \bibinfo{person}{Klaus Ostermann}.}
  \bibinfo{year}{2016}\natexlab{}.
\newblock \showarticletitle{Parsing with First-class Derivatives}. In
  \bibinfo{booktitle}{\emph{Proceedings of the 2016 ACM SIGPLAN International
  Conference on Object-Oriented Programming, Systems, Languages, and
  Applications}} (Amsterdam, Netherlands) \emph{(\bibinfo{series}{OOPSLA
  2016})}. \bibinfo{publisher}{ACM}, \bibinfo{address}{New York, NY, USA},
  \bibinfo{pages}{588--606}.
\newblock
\showISBNx{978-1-4503-4444-9}
\urldef\tempurl%
\url{https://doi.org/10.1145/2983990.2984026}
\showDOI{\tempurl}


\bibitem[\protect\citeauthoryear{Br{\"u}ggemann-Klein and
  Wood}{Br{\"u}ggemann-Klein and Wood}{1992}]%
        {bruggemann1992deterministic}
\bibfield{author}{\bibinfo{person}{Anne Br{\"u}ggemann-Klein} {and}
  \bibinfo{person}{Derick Wood}.} \bibinfo{year}{1992}\natexlab{}.
\newblock \showarticletitle{Deterministic regular languages}. In
  \bibinfo{booktitle}{\emph{Annual Symposium on Theoretical Aspects of Computer
  Science}}. Springer, \bibinfo{pages}{173--184}.
\newblock


\bibitem[\protect\citeauthoryear{Brzozowski}{Brzozowski}{1964}]%
        {brzozowski1964derivatives}
\bibfield{author}{\bibinfo{person}{Janusz~A Brzozowski}.}
  \bibinfo{year}{1964}\natexlab{}.
\newblock \showarticletitle{Derivatives of regular expressions}. In
  \bibinfo{booktitle}{\emph{Journal of the ACM}}. Citeseer.
\newblock


\bibitem[\protect\citeauthoryear{Burge}{Burge}{1975}]%
        {burge1975recursive}
\bibfield{author}{\bibinfo{person}{William~H Burge}.}
  \bibinfo{year}{1975}\natexlab{}.
\newblock \showarticletitle{Recursive programming techniques}.
\newblock  (\bibinfo{year}{1975}).
\newblock


\bibitem[\protect\citeauthoryear{Cocke}{Cocke}{1969}]%
        {cocke1969programming}
\bibfield{author}{\bibinfo{person}{John Cocke}.}
  \bibinfo{year}{1969}\natexlab{}.
\newblock \showarticletitle{Programming languages and their compilers:
  Preliminary notes}.
\newblock  (\bibinfo{year}{1969}).
\newblock


\bibitem[\protect\citeauthoryear{Danielsson}{Danielsson}{2010}]%
        {Danielsson:2010:TPC:1863543.1863585}
\bibfield{author}{\bibinfo{person}{Nils~Anders Danielsson}.}
  \bibinfo{year}{2010}\natexlab{}.
\newblock \showarticletitle{Total Parser Combinators}. In
  \bibinfo{booktitle}{\emph{Proceedings of the 15th ACM SIGPLAN International
  Conference on Functional Programming}} (Baltimore, Maryland, USA)
  \emph{(\bibinfo{series}{ICFP '10})}. \bibinfo{publisher}{ACM},
  \bibinfo{address}{New York, NY, USA}, \bibinfo{pages}{285--296}.
\newblock
\showISBNx{978-1-60558-794-3}
\urldef\tempurl%
\url{https://doi.org/10.1145/1863543.1863585}
\showDOI{\tempurl}


\bibitem[\protect\citeauthoryear{DeRemer}{DeRemer}{1969}]%
        {deremer1969practical}
\bibfield{author}{\bibinfo{person}{Franklin~Lewis DeRemer}.}
  \bibinfo{year}{1969}\natexlab{}.
\newblock \emph{\bibinfo{title}{Practical translators for LR (k) languages.}}
\newblock \bibinfo{thesistype}{Ph.D. Dissertation}.
  \bibinfo{school}{Massachusetts Institute of Technology}.
\newblock


\bibitem[\protect\citeauthoryear{Earley}{Earley}{1970}]%
        {earley1970efficient}
\bibfield{author}{\bibinfo{person}{Jay Earley}.}
  \bibinfo{year}{1970}\natexlab{}.
\newblock \showarticletitle{An efficient context-free parsing algorithm}.
\newblock \bibinfo{journal}{\emph{Commun. ACM}} \bibinfo{volume}{13},
  \bibinfo{number}{2} (\bibinfo{year}{1970}), \bibinfo{pages}{94--102}.
\newblock


\bibitem[\protect\citeauthoryear{Fokker}{Fokker}{1995}]%
        {fokker1995functional}
\bibfield{author}{\bibinfo{person}{Jeroen Fokker}.}
  \bibinfo{year}{1995}\natexlab{}.
\newblock \showarticletitle{Functional parsers}. In
  \bibinfo{booktitle}{\emph{International School on Advanced Functional
  Programming}}. Springer, \bibinfo{pages}{1--23}.
\newblock


\bibitem[\protect\citeauthoryear{Ford}{Ford}{2002}]%
        {ford2002packrat}
\bibfield{author}{\bibinfo{person}{Bryan Ford}.}
  \bibinfo{year}{2002}\natexlab{}.
\newblock \showarticletitle{Packrat Parsing:: Simple, Powerful, Lazy, Linear
  Time, Functional Pearl}. In \bibinfo{booktitle}{\emph{Proceedings of the
  Seventh ACM SIGPLAN International Conference on Functional Programming}}
  (Pittsburgh, PA, USA) \emph{(\bibinfo{series}{ICFP '02})}.
  \bibinfo{publisher}{ACM}, \bibinfo{address}{New York, NY, USA},
  \bibinfo{pages}{36--47}.
\newblock
\showISBNx{1-58113-487-8}
\urldef\tempurl%
\url{https://doi.org/10.1145/581478.581483}
\showDOI{\tempurl}


\bibitem[\protect\citeauthoryear{Ford}{Ford}{2004}]%
        {ford2004parsing}
\bibfield{author}{\bibinfo{person}{Bryan Ford}.}
  \bibinfo{year}{2004}\natexlab{}.
\newblock \showarticletitle{Parsing Expression Grammars: A Recognition-based
  Syntactic Foundation}. In \bibinfo{booktitle}{\emph{Proceedings of the 31st
  ACM SIGPLAN-SIGACT Symposium on Principles of Programming Languages}}
  (Venice, Italy) \emph{(\bibinfo{series}{POPL '04})}.
  \bibinfo{publisher}{ACM}, \bibinfo{address}{New York, NY, USA},
  \bibinfo{pages}{111--122}.
\newblock
\showISBNx{1-58113-729-X}
\urldef\tempurl%
\url{https://doi.org/10.1145/964001.964011}
\showDOI{\tempurl}


\bibitem[\protect\citeauthoryear{Grimm}{Grimm}{2004}]%
        {grimm2004practical}
\bibfield{author}{\bibinfo{person}{Robert Grimm}.}
  \bibinfo{year}{2004}\natexlab{}.
\newblock \bibinfo{booktitle}{\emph{Practical Packrat Parsing}}.
\newblock \bibinfo{type}{{T}echnical {R}eport}. \bibinfo{institution}{New York
  University}.
\newblock


\bibitem[\protect\citeauthoryear{Haoyi}{Haoyi}{2019}]%
        {Haoyi19FastParse}
\bibfield{author}{\bibinfo{person}{Li Haoyi}.} \bibinfo{year}{2019}\natexlab{}.
\newblock \bibinfo{title}{FastParse 2.1.3}.
\newblock \bibinfo{howpublished}{\url{http://www.lihaoyi.com/fastparse/}}.
\newblock


\bibitem[\protect\citeauthoryear{Henriksen, Bilardi, and Pingali}{Henriksen
  et~al\mbox{.}}{2019}]%
        {Henriksen:2019:DGS:3366395.3360553}
\bibfield{author}{\bibinfo{person}{Ian Henriksen}, \bibinfo{person}{Gianfranco
  Bilardi}, {and} \bibinfo{person}{Keshav Pingali}.}
  \bibinfo{year}{2019}\natexlab{}.
\newblock \showarticletitle{Derivative Grammars: A Symbolic Approach to Parsing
  with Derivatives}.
\newblock \bibinfo{journal}{\emph{Proc. ACM Program. Lang.}}
  \bibinfo{volume}{3}, \bibinfo{number}{OOPSLA}, Article
  \bibinfo{articleno}{127} (\bibinfo{date}{Oct.} \bibinfo{year}{2019}),
  \bibinfo{numpages}{28}~pages.
\newblock
\showISSN{2475-1421}
\urldef\tempurl%
\url{https://doi.org/10.1145/3360553}
\showDOI{\tempurl}


\bibitem[\protect\citeauthoryear{Huet}{Huet}{1997}]%
        {huet1997zipper}
\bibfield{author}{\bibinfo{person}{G{\'e}rard Huet}.}
  \bibinfo{year}{1997}\natexlab{}.
\newblock \showarticletitle{The zipper}.
\newblock \bibinfo{journal}{\emph{Journal of functional programming}}
  \bibinfo{volume}{7}, \bibinfo{number}{5} (\bibinfo{year}{1997}),
  \bibinfo{pages}{549--554}.
\newblock


\bibitem[\protect\citeauthoryear{Hutton}{Hutton}{1992}]%
        {hutton1992higher}
\bibfield{author}{\bibinfo{person}{Graham Hutton}.}
  \bibinfo{year}{1992}\natexlab{}.
\newblock \showarticletitle{Higher-order functions for parsing}.
\newblock \bibinfo{journal}{\emph{Journal of functional programming}}
  \bibinfo{volume}{2}, \bibinfo{number}{3} (\bibinfo{year}{1992}),
  \bibinfo{pages}{323--343}.
\newblock


\bibitem[\protect\citeauthoryear{Hutton and Meijer}{Hutton and Meijer}{1996}]%
        {hutton1996monadic}
\bibfield{author}{\bibinfo{person}{Graham Hutton} {and} \bibinfo{person}{Erik
  Meijer}.} \bibinfo{year}{1996}\natexlab{}.
\newblock \showarticletitle{Monadic parser combinators}.
\newblock  (\bibinfo{year}{1996}).
\newblock


\bibitem[\protect\citeauthoryear{Johnstone and Scott}{Johnstone and
  Scott}{1998}]%
        {johnstone1998generalised}
\bibfield{author}{\bibinfo{person}{Adrian Johnstone} {and}
  \bibinfo{person}{Elizabeth Scott}.} \bibinfo{year}{1998}\natexlab{}.
\newblock \showarticletitle{Generalised recursive descent parsing and
  follow-determinism}. In \bibinfo{booktitle}{\emph{International Conference on
  Compiler Construction}}. Springer, \bibinfo{pages}{16--30}.
\newblock


\bibitem[\protect\citeauthoryear{Jourdan, Pottier, and Leroy}{Jourdan
  et~al\mbox{.}}{2012}]%
        {jourdan2012validating}
\bibfield{author}{\bibinfo{person}{Jacques-Henri Jourdan},
  \bibinfo{person}{Fran{\c{c}}ois Pottier}, {and} \bibinfo{person}{Xavier
  Leroy}.} \bibinfo{year}{2012}\natexlab{}.
\newblock \showarticletitle{Validating LR (1) parsers}. In
  \bibinfo{booktitle}{\emph{European Symposium on Programming}}. Springer,
  \bibinfo{pages}{397--416}.
\newblock


\bibitem[\protect\citeauthoryear{Kasami}{Kasami}{1966}]%
        {kasami1966efficient}
\bibfield{author}{\bibinfo{person}{Tadao Kasami}.}
  \bibinfo{year}{1966}\natexlab{}.
\newblock \showarticletitle{An efficient recognition and syntax-analysis
  algorithm for context-free languages}.
\newblock \bibinfo{journal}{\emph{Coordinated Science Laboratory Report no.
  R-257}} (\bibinfo{year}{1966}).
\newblock


\bibitem[\protect\citeauthoryear{Knuth}{Knuth}{1965}]%
        {knuth1965translation}
\bibfield{author}{\bibinfo{person}{Donald~E Knuth}.}
  \bibinfo{year}{1965}\natexlab{}.
\newblock \showarticletitle{On the translation of languages from left to
  right}.
\newblock \bibinfo{journal}{\emph{Information and control}}
  \bibinfo{volume}{8}, \bibinfo{number}{6} (\bibinfo{year}{1965}),
  \bibinfo{pages}{607--639}.
\newblock


\bibitem[\protect\citeauthoryear{Koprowski and Binsztok}{Koprowski and
  Binsztok}{2010}]%
        {koprowski2010trx}
\bibfield{author}{\bibinfo{person}{Adam Koprowski} {and} \bibinfo{person}{Henri
  Binsztok}.} \bibinfo{year}{2010}\natexlab{}.
\newblock \showarticletitle{TRX: A formally verified parser interpreter}. In
  \bibinfo{booktitle}{\emph{European Symposium on Programming}}. Springer,
  \bibinfo{pages}{345--365}.
\newblock


\bibitem[\protect\citeauthoryear{Krishnaswami and Yallop}{Krishnaswami and
  Yallop}{2019}]%
        {Krishnaswami:2019:TAA:3314221.3314625}
\bibfield{author}{\bibinfo{person}{Neelakantan~R. Krishnaswami} {and}
  \bibinfo{person}{Jeremy Yallop}.} \bibinfo{year}{2019}\natexlab{}.
\newblock \showarticletitle{A Typed, Algebraic Approach to Parsing}. In
  \bibinfo{booktitle}{\emph{Proceedings of the 40th ACM SIGPLAN Conference on
  Programming Language Design and Implementation}} (Phoenix, AZ, USA)
  \emph{(\bibinfo{series}{PLDI 2019})}. \bibinfo{publisher}{ACM},
  \bibinfo{address}{New York, NY, USA}, \bibinfo{pages}{379--393}.
\newblock
\showISBNx{978-1-4503-6712-7}
\urldef\tempurl%
\url{https://doi.org/10.1145/3314221.3314625}
\showDOI{\tempurl}


\bibitem[\protect\citeauthoryear{Kumar, Myreen, Norrish, and Owens}{Kumar
  et~al\mbox{.}}{2014}]%
        {kumar2014cakeml}
\bibfield{author}{\bibinfo{person}{Ramana Kumar}, \bibinfo{person}{Magnus~O.
  Myreen}, \bibinfo{person}{Michael Norrish}, {and} \bibinfo{person}{Scott
  Owens}.} \bibinfo{year}{2014}\natexlab{}.
\newblock \showarticletitle{CakeML: A Verified Implementation of ML}. In
  \bibinfo{booktitle}{\emph{Proceedings of the 41st ACM SIGPLAN-SIGACT
  Symposium on Principles of Programming Languages}} (San Diego, California,
  USA) \emph{(\bibinfo{series}{POPL '14})}. \bibinfo{publisher}{ACM},
  \bibinfo{address}{New York, NY, USA}, \bibinfo{pages}{179--191}.
\newblock
\showISBNx{978-1-4503-2544-8}
\urldef\tempurl%
\url{https://doi.org/10.1145/2535838.2535841}
\showDOI{\tempurl}


\bibitem[\protect\citeauthoryear{{LAMP EPFL} and {Lightbend, Inc}}{{LAMP EPFL}
  and {Lightbend, Inc}}{2019}]%
        {scalaparsercombinators}
\bibfield{author}{\bibinfo{person}{{LAMP EPFL}} {and}
  \bibinfo{person}{{Lightbend, Inc}}.} \bibinfo{year}{2019}\natexlab{}.
\newblock \bibinfo{title}{Scala Parser Combinators}.
\newblock
  \bibinfo{howpublished}{\url{https://github.com/scala/scala-parser-combinators}}.
\newblock


\bibitem[\protect\citeauthoryear{Lang}{Lang}{1974}]%
        {lang1974deterministic}
\bibfield{author}{\bibinfo{person}{Bernard Lang}.}
  \bibinfo{year}{1974}\natexlab{}.
\newblock \showarticletitle{Deterministic techniques for efficient
  non-deterministic parsers}. In \bibinfo{booktitle}{\emph{International
  Colloquium on Automata, Languages, and Programming}}. Springer,
  \bibinfo{pages}{255--269}.
\newblock


\bibitem[\protect\citeauthoryear{Lasser, Casinghino, Fisher, and Roux}{Lasser
  et~al\mbox{.}}{2019}]%
        {lasser2019verified}
\bibfield{author}{\bibinfo{person}{Sam Lasser}, \bibinfo{person}{Chris
  Casinghino}, \bibinfo{person}{Kathleen Fisher}, {and} \bibinfo{person}{Cody
  Roux}.} \bibinfo{year}{2019}\natexlab{}.
\newblock \showarticletitle{A Verified LL (1) Parser Generator}. In
  \bibinfo{booktitle}{\emph{10th International Conference on Interactive
  Theorem Proving (ITP 2019)}}. Schloss Dagstuhl-Leibniz-Zentrum fuer
  Informatik.
\newblock


\bibitem[\protect\citeauthoryear{Leijen and Meijer}{Leijen and Meijer}{2001}]%
        {leijen2001parsec}
\bibfield{author}{\bibinfo{person}{Daan Leijen} {and} \bibinfo{person}{Erik
  Meijer}.} \bibinfo{year}{2001}\natexlab{}.
\newblock \showarticletitle{Parsec: Direct style monadic parser combinators for
  the real world}.
\newblock  (\bibinfo{year}{2001}).
\newblock


\bibitem[\protect\citeauthoryear{Lei{\ss}}{Lei{\ss}}{1991}]%
        {leiss1991towards}
\bibfield{author}{\bibinfo{person}{Haas Lei{\ss}}.}
  \bibinfo{year}{1991}\natexlab{}.
\newblock \showarticletitle{Towards Kleene algebra with recursion}. In
  \bibinfo{booktitle}{\emph{International Workshop on Computer Science Logic}}.
  Springer, \bibinfo{pages}{242--256}.
\newblock


\bibitem[\protect\citeauthoryear{Leroy}{Leroy}{2009}]%
        {leroy2009formal}
\bibfield{author}{\bibinfo{person}{Xavier Leroy}.}
  \bibinfo{year}{2009}\natexlab{}.
\newblock \showarticletitle{Formal verification of a realistic compiler}.
\newblock \bibinfo{journal}{\emph{Commun. ACM}} \bibinfo{volume}{52},
  \bibinfo{number}{7} (\bibinfo{year}{2009}), \bibinfo{pages}{107--115}.
\newblock


\bibitem[\protect\citeauthoryear{Lewis and Stearns}{Lewis and Stearns}{1968}]%
        {Lewis:1968:ST:321466.321477}
\bibfield{author}{\bibinfo{person}{P.~M. Lewis, II} {and}
  \bibinfo{person}{R.~E. Stearns}.} \bibinfo{year}{1968}\natexlab{}.
\newblock \showarticletitle{Syntax-Directed Transduction}.
\newblock \bibinfo{journal}{\emph{J. ACM}} \bibinfo{volume}{15},
  \bibinfo{number}{3} (\bibinfo{date}{July} \bibinfo{year}{1968}),
  \bibinfo{pages}{465--488}.
\newblock
\showISSN{0004-5411}
\urldef\tempurl%
\url{https://doi.org/10.1145/321466.321477}
\showDOI{\tempurl}


\bibitem[\protect\citeauthoryear{McBride}{McBride}{2001}]%
        {Mcbride01thederivative}
\bibfield{author}{\bibinfo{person}{Conor McBride}.}
  \bibinfo{year}{2001}\natexlab{}.
\newblock \bibinfo{title}{The Derivative of a Regular Type is its Type of
  One-Hole Contexts (Extended Abstract)}.
\newblock
\newblock


\bibitem[\protect\citeauthoryear{McBride and Paterson}{McBride and
  Paterson}{2008}]%
        {mcbride2008applicative}
\bibfield{author}{\bibinfo{person}{Conor McBride} {and} \bibinfo{person}{Ross
  Paterson}.} \bibinfo{year}{2008}\natexlab{}.
\newblock \showarticletitle{Applicative programming with effects}.
\newblock \bibinfo{journal}{\emph{Journal of functional programming}}
  \bibinfo{volume}{18}, \bibinfo{number}{1} (\bibinfo{year}{2008}),
  \bibinfo{pages}{1--13}.
\newblock


\bibitem[\protect\citeauthoryear{Might, Darais, and Spiewak}{Might
  et~al\mbox{.}}{2011}]%
        {might2011parsing}
\bibfield{author}{\bibinfo{person}{Matthew Might}, \bibinfo{person}{David
  Darais}, {and} \bibinfo{person}{Daniel Spiewak}.}
  \bibinfo{year}{2011}\natexlab{}.
\newblock \showarticletitle{Parsing with Derivatives: A Functional Pearl}. In
  \bibinfo{booktitle}{\emph{Proceedings of the 16th ACM SIGPLAN International
  Conference on Functional Programming}} (Tokyo, Japan)
  \emph{(\bibinfo{series}{ICFP '11})}. \bibinfo{publisher}{ACM},
  \bibinfo{address}{New York, NY, USA}, \bibinfo{pages}{189--195}.
\newblock
\showISBNx{978-1-4503-0865-6}
\urldef\tempurl%
\url{https://doi.org/10.1145/2034773.2034801}
\showDOI{\tempurl}


\bibitem[\protect\citeauthoryear{Omanashvili}{Omanashvili}{2019}]%
        {JSONGenerator}
\bibfield{author}{\bibinfo{person}{Vazha Omanashvili}.}
  \bibinfo{year}{2019}\natexlab{}.
\newblock \bibinfo{title}{JSON Generator}.
\newblock \bibinfo{howpublished}{\url{https://www.json-generator.com}}.
\newblock
\newblock
\shownote{Accessed 2019-11-20.}


\bibitem[\protect\citeauthoryear{Parr}{Parr}{2013}]%
        {parr2013definitive}
\bibfield{author}{\bibinfo{person}{Terence Parr}.}
  \bibinfo{year}{2013}\natexlab{}.
\newblock \bibinfo{booktitle}{\emph{The definitive ANTLR 4 reference}}.
\newblock \bibinfo{publisher}{Pragmatic Bookshelf}.
\newblock


\bibitem[\protect\citeauthoryear{Parr}{Parr}{2019}]%
        {ANTLRexamples}
\bibfield{author}{\bibinfo{person}{Terence Parr}.}
  \bibinfo{year}{2019}\natexlab{}.
\newblock \bibinfo{title}{Grammars written for ANTLR v4; expectation that the
  grammars are free of actions}.
\newblock
  \bibinfo{howpublished}{\url{https://github.com/antlr/grammars-v4/tree/master/json}}.
\newblock
\newblock
\shownote{Accessed 2019-11-22.}


\bibitem[\protect\citeauthoryear{Parr and Fisher}{Parr and Fisher}{2011}]%
        {DBLP:conf/pldi/ParrF11}
\bibfield{author}{\bibinfo{person}{Terence Parr} {and}
  \bibinfo{person}{Kathleen Fisher}.} \bibinfo{year}{2011}\natexlab{}.
\newblock \showarticletitle{LL(*): the foundation of the {ANTLR} parser
  generator}. In \bibinfo{booktitle}{\emph{Proceedings of the 32nd {ACM}
  {SIGPLAN} Conference on Programming Language Design and Implementation,
  {PLDI} 2011, San Jose, CA, USA, June 4-8, 2011}}. \bibinfo{pages}{425--436}.
\newblock
\urldef\tempurl%
\url{https://doi.org/10.1145/1993498.1993548}
\showDOI{\tempurl}


\bibitem[\protect\citeauthoryear{Pierce, de~Amorim, Casinghino, Gaboardi,
  Greenberg, Hri{\c t}cu, Sj{\"o}berg, and Yorgey}{Pierce
  et~al\mbox{.}}{2018}]%
        {Pierce:SF1}
\bibfield{author}{\bibinfo{person}{Benjamin~C. Pierce},
  \bibinfo{person}{Arthur~Azevedo de Amorim}, \bibinfo{person}{Chris
  Casinghino}, \bibinfo{person}{Marco Gaboardi}, \bibinfo{person}{Michael
  Greenberg}, \bibinfo{person}{C{\u a}t{\u a}lin Hri{\c t}cu},
  \bibinfo{person}{Vilhelm Sj{\"o}berg}, {and} \bibinfo{person}{Brent Yorgey}.}
  \bibinfo{year}{2018}\natexlab{}.
\newblock \bibinfo{booktitle}{\emph{Logical Foundations}}.
\newblock \bibinfo{publisher}{Electronic textbook}.
\newblock
\newblock
\shownote{Version 5.5. \url{http://www.cis.upenn.edu/~bcpierce/sf}.}


\bibitem[\protect\citeauthoryear{Prokopec}{Prokopec}{2019}]%
        {Scalameter}
\bibfield{author}{\bibinfo{person}{Aleksandar Prokopec}.}
  \bibinfo{year}{2019}\natexlab{}.
\newblock \bibinfo{title}{Scalameter: Automate your performance testing today}.
\newblock \bibinfo{howpublished}{\url{https://scalameter.github.io}/}.
\newblock
\newblock
\shownote{Accessed 2019-11-20.}


\bibitem[\protect\citeauthoryear{Radul}{Radul}{2009}]%
        {radul2009propagation}
\bibfield{author}{\bibinfo{person}{Alexey Radul}.}
  \bibinfo{year}{2009}\natexlab{}.
\newblock \showarticletitle{Propagation networks: A flexible and expressive
  substrate for computation}.
\newblock  (\bibinfo{year}{2009}).
\newblock


\bibitem[\protect\citeauthoryear{Ramananandro, Delignat{-}Lavaud, Fournet,
  Swamy, Chajed, Kobeissi, and Protzenko}{Ramananandro et~al\mbox{.}}{2019}]%
        {DBLP:conf/uss/RamananandroDFS19}
\bibfield{author}{\bibinfo{person}{Tahina Ramananandro},
  \bibinfo{person}{Antoine Delignat{-}Lavaud}, \bibinfo{person}{C{\'{e}}dric
  Fournet}, \bibinfo{person}{Nikhil Swamy}, \bibinfo{person}{Tej Chajed},
  \bibinfo{person}{Nadim Kobeissi}, {and} \bibinfo{person}{Jonathan
  Protzenko}.} \bibinfo{year}{2019}\natexlab{}.
\newblock \showarticletitle{EverParse: Verified Secure Zero-Copy Parsers for
  Authenticated Message Formats}. In \bibinfo{booktitle}{\emph{28th {USENIX}
  Security Symposium, {USENIX} Security 2019, Santa Clara, CA, USA, August
  14-16, 2019}}. \bibinfo{pages}{1465--1482}.
\newblock
\urldef\tempurl%
\url{https://www.usenix.org/conference/usenixsecurity19/presentation/delignat-lavaud}
\showURL{%
\tempurl}


\bibitem[\protect\citeauthoryear{Redziejowski}{Redziejowski}{2008}]%
        {redziejowski2008some}
\bibfield{author}{\bibinfo{person}{Roman~R Redziejowski}.}
  \bibinfo{year}{2008}\natexlab{}.
\newblock \showarticletitle{Some aspects of parsing expression grammar}.
\newblock \bibinfo{journal}{\emph{Fundamenta Informaticae}}
  \bibinfo{volume}{85}, \bibinfo{number}{1-4} (\bibinfo{year}{2008}),
  \bibinfo{pages}{441--451}.
\newblock


\bibitem[\protect\citeauthoryear{Rendel and Ostermann}{Rendel and
  Ostermann}{2010}]%
        {rendel2010invertible}
\bibfield{author}{\bibinfo{person}{Tillmann Rendel} {and}
  \bibinfo{person}{Klaus Ostermann}.} \bibinfo{year}{2010}\natexlab{}.
\newblock \showarticletitle{Invertible Syntax Descriptions: Unifying Parsing
  and Pretty Printing}. In \bibinfo{booktitle}{\emph{Proceedings of the Third
  ACM Haskell Symposium on Haskell}} (Baltimore, Maryland, USA)
  \emph{(\bibinfo{series}{Haskell '10})}. \bibinfo{publisher}{ACM},
  \bibinfo{address}{New York, NY, USA}, \bibinfo{pages}{1--12}.
\newblock
\showISBNx{978-1-4503-0252-4}
\urldef\tempurl%
\url{https://doi.org/10.1145/1863523.1863525}
\showDOI{\tempurl}


\bibitem[\protect\citeauthoryear{Scott and Johnstone}{Scott and
  Johnstone}{2010}]%
        {scott2010gll}
\bibfield{author}{\bibinfo{person}{Elizabeth Scott} {and}
  \bibinfo{person}{Adrian Johnstone}.} \bibinfo{year}{2010}\natexlab{}.
\newblock \showarticletitle{GLL parsing}.
\newblock \bibinfo{journal}{\emph{Electronic Notes in Theoretical Computer
  Science}} \bibinfo{volume}{253}, \bibinfo{number}{7} (\bibinfo{year}{2010}),
  \bibinfo{pages}{177--189}.
\newblock


\bibitem[\protect\citeauthoryear{Sozeau and Mangin}{Sozeau and Mangin}{2019}]%
        {sozeau2019equations}
\bibfield{author}{\bibinfo{person}{Matthieu Sozeau} {and}
  \bibinfo{person}{Cyprien Mangin}.} \bibinfo{year}{2019}\natexlab{}.
\newblock \showarticletitle{Equations reloaded: high-level dependently-typed
  functional programming and proving in Coq}.
\newblock \bibinfo{journal}{\emph{Proceedings of the ACM on Programming
  Languages}} \bibinfo{volume}{3}, \bibinfo{number}{ICFP}
  (\bibinfo{year}{2019}), \bibinfo{pages}{86}.
\newblock


\bibitem[\protect\citeauthoryear{Spiewak}{Spiewak}{2018}]%
        {parseback}
\bibfield{author}{\bibinfo{person}{Daniel Spiewak}.}
  \bibinfo{year}{2018}\natexlab{}.
\newblock \bibinfo{title}{Parseback}.
\newblock \bibinfo{howpublished}{\url{https://github.com/djspiewak/parseback}}.
\newblock


\bibitem[\protect\citeauthoryear{Steele~Jr}{Steele~Jr}{1980}]%
        {steele1980definition}
\bibfield{author}{\bibinfo{person}{Guy~L Steele~Jr}.}
  \bibinfo{year}{1980}\natexlab{}.
\newblock \showarticletitle{The definition and implementation of a computer
  programming language based on constraints}.
\newblock  (\bibinfo{year}{1980}).
\newblock


\bibitem[\protect\citeauthoryear{Swierstra and Duponcheel}{Swierstra and
  Duponcheel}{1996}]%
        {swierstra1996deterministic}
\bibfield{author}{\bibinfo{person}{S~Doaitse Swierstra} {and}
  \bibinfo{person}{Luc Duponcheel}.} \bibinfo{year}{1996}\natexlab{}.
\newblock \showarticletitle{Deterministic, error-correcting combinator
  parsers}. In \bibinfo{booktitle}{\emph{International School on Advanced
  Functional Programming}}. Springer, \bibinfo{pages}{184--207}.
\newblock


\bibitem[\protect\citeauthoryear{Traytel}{Traytel}{2015}]%
        {Formula_Derivatives-AFP}
\bibfield{author}{\bibinfo{person}{Dmitriy Traytel}.}
  \bibinfo{year}{2015}\natexlab{}.
\newblock \showarticletitle{Derivatives of Logical Formulas}.
\newblock \bibinfo{journal}{\emph{Archive of Formal Proofs}}
  (\bibinfo{date}{May} \bibinfo{year}{2015}).
\newblock
\showISSN{2150-914x}
\newblock
\shownote{\url{http://isa-afp.org/entries/Formula_Derivatives.html}, Formal
  proof development.}


\bibitem[\protect\citeauthoryear{Traytel and Nipkow}{Traytel and
  Nipkow}{2014}]%
        {MSO_Regex_Equivalence-AFP}
\bibfield{author}{\bibinfo{person}{Dmitriy Traytel} {and}
  \bibinfo{person}{Tobias Nipkow}.} \bibinfo{year}{2014}\natexlab{}.
\newblock \showarticletitle{Decision Procedures for MSO on Words Based on
  Derivatives of Regular Expressions}.
\newblock \bibinfo{journal}{\emph{Archive of Formal Proofs}}
  (\bibinfo{date}{June} \bibinfo{year}{2014}).
\newblock
\showISSN{2150-914x}
\newblock
\shownote{\url{http://isa-afp.org/entries/MSO_Regex_Equivalence.html}, Formal
  proof development.}


\bibitem[\protect\citeauthoryear{Wadler}{Wadler}{1985}]%
        {wadler1985replace}
\bibfield{author}{\bibinfo{person}{Philip Wadler}.}
  \bibinfo{year}{1985}\natexlab{}.
\newblock \showarticletitle{How to replace failure by a list of successes a
  method for exception handling, backtracking, and pattern matching in lazy
  functional languages}. In \bibinfo{booktitle}{\emph{Conference on Functional
  Programming Languages and Computer Architecture}}. Springer,
  \bibinfo{pages}{113--128}.
\newblock


\bibitem[\protect\citeauthoryear{Younger}{Younger}{1967}]%
        {younger1967recognition}
\bibfield{author}{\bibinfo{person}{Daniel~H Younger}.}
  \bibinfo{year}{1967}\natexlab{}.
\newblock \showarticletitle{Recognition and parsing of context-free languages
  in time $n^3$}.
\newblock \bibinfo{journal}{\emph{Information and control}}
  \bibinfo{volume}{10}, \bibinfo{number}{2} (\bibinfo{year}{1967}),
  \bibinfo{pages}{189--208}.
\newblock


\end{thebibliography}
